\documentclass[AMA,Times1COL]{WileyNJDv5} 

\articletype{RESEARCH ARTICLE}%

\journal{Statistics in Medicine}
\usepackage{pdfpages}

\begin{document}

\title{A modified debiased inverse-variance weighted estimator in two-sample summary-data Mendelian randomization}

\author[1]{Youpeng Su}
\author[2]{Siqi Xu}
\author[1]{Yilei Ma}
\author[1]{Ping Yin}
\author[2] {Wing Kam Fung}
\author[1]{Hongwei Jiang}
\author[1]{Peng Wang}

\authormark{SU \textsc{et al.}}
\titlemark{c}

\address[1]{\orgdiv{Department of Epidemiology and Biostatistics, School of Public Health, Tongji Medical College}, 
	\orgname{Huazhong University of Science and Technology}, \orgaddress{\state{Wuhan}, \country{China}}}
\address[2]{\orgdiv{Department of Statistics and Actuarial Science}, 
	\orgname{The University of Hong Kong}, \orgaddress{\state{Hong Kong SAR}, \country{China}}}
	
\corres{Peng Wang, Department of Epidemiology and Biostatistics, School of Public Health, Tongji Medical College, Huazhong University of Science and Technology, Hangkong Road 13, Wuhan, China.
	\email{pengwang\_stat@hust.edu.cn} \\
	Hongwei Jiang, Department of Epidemiology and Biostatistics, School of Public Health, Tongji Medical College, Huazhong University of Science and Technology, Hangkong Road 13, Wuhan, China.
	\email{jhwccc0@sina.com}}


\fundingInfo{National Natural Science Foundation of China, Grant/Award Numbers 82173628, 82173619; 
	the Fundamental Research Funds for School of Public Health, Tongji Medical College, Huazhong University of Science and Technology, Grant/Award 2022gwzz03}

\abstract[Abstract]{Mendelian randomization uses genetic variants as instrumental variables to make
	causal inferences about the effects of modifiable risk factors on diseases from observational
	data. One of the major challenges in Mendelian randomization is that many
	genetic variants are only modestly or even weakly associated with the risk factor of
	interest, a setting known as many weak instruments. Many existing methods, such
	as the popular inverse-variance weighted (IVW) method, could be biased when the
	instrument strength is weak. To address this issue, the debiased IVW (dIVW) estimator,
	which is shown to be robust to many weak instruments, was recently proposed.
	However, this estimator still has non-ignorable bias when the effective sample size
	is small. In this paper, we propose a modified debiased IVW (mdIVW) estimator by
	multiplying a modification factor to the original dIVW estimator. After this simple
	correction, we show that the bias of the mdIVW estimator converges to zero at a
	faster rate than that of the dIVW estimator under some regularity conditions. Moreover,
	the mdIVW estimator has smaller variance than the dIVW estimator.We further
	extend the proposed method to account for the presence of instrumental variable selection
	and balanced horizontal pleiotropy. We demonstrate the improvement of the
	mdIVW estimator over the dIVW estimator through extensive simulation studies and
	real data analysis.
}

\maketitle

\section{Introduction}\label{sec1}
\begin{figure}[htb]
	\centering
	\includegraphics[width=0.8\textwidth]{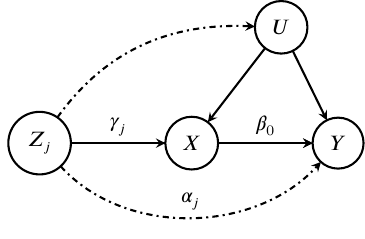}
	\caption{Illustrative diagram shows the core IV assumptions for the 
		genetic variant $Z_{j}$ (solid lines) with potential violations (dashed lines).
		$U$ is the confounding.
		$\gamma_{j}$ is the genetic effect of $Z_j$ on the exposure $X$.
		$\alpha_{j}$ is  the direct genetic effect of $Z_j$ on the outcome $Y$.
		And $\beta_{0}$ is the causal effect of the exposure $X$ on the outcome $Y$.}
	\label{fig1}
\end{figure}

A common goal in epidemiology is to study the causal effects of modifiable risk factors on health outcomes. If a risk
factor is known to impair health causally, interventions to eliminate or reduce exposure to that risk factor can be implemented
to promote the population health. Although the randomized controlled trial (RCT) provides the highest quality of
evidence of causality, most inferences about causality are drawn from observational data, which are more readily available.
However, the statistical results given by observational studies could be biased if there exists unmeasured confounding.\cite{Zhao2020,San2007}
To address this issue, Mendelian randomization (MR) uses genetic variants, usually single nucleotide polymorphisms (SNPs), as instrumental variables (IVs) to generate an unbiased estimate of the causal effect even in the presence of unmeasured
confounding.\cite{Burgess2017} Nonetheless, for a SNP to be a valid IV, it must satisfy three well-known core assumptions (see
Figure \ref{fig1} for a graphical illustration):\cite{Didelez2007}
\begin{enumerate}[1.]
	\item IV relevance: the SNP must be associated with the exposure;
	\item IV independence: the SNP is independent of any confounders of the exposure-outcome relationship;
	\item Exclusion restriction: the SNP affects the outcome only through the exposure. 
\end{enumerate}

A major challenge in MR is that many genetic variants are only modestly or weakly associated with the risk factor of
interest, a setting known as many weak instruments, where the genetic variants explain only a small fraction of the variance
of the exposure.\cite{Davies2015,Chao2005} In this case, the IV relevance assumption is nearly violated, and as a result, the weak IV bias is likely
to be introduced.\cite{Bound1995} The violation of the IV independence assumption and the exclusion restriction assumption due to the
widespread horizontal pleiotropy is another concern where the genetic variants directly affect the outcome besides being
mediated by the exposure variable.\cite{Hemani2018,Verbanck2018} If any of these three core assumptions is violated, conventional MR methods may
produce biased estimates. Some MR methods for individual-level data are robust to these violations.\cite{Tchetgen2021,Pacini2016,Guo2018,Kang2016} However, the
access to individual-level data is rather restricted due to privacy concerns and various logistical considerations. Instead,
extensive summary statistics estimated from genome-wide association studies (GWASs) are publicly available. Therefore,
many recently proposed MR methods were adapted to use GWAS summary statistics as input.\cite{Sanderson2022}
In this paper, we focus
on a popular setting in MR known as two-sample summary-data MR, where the summary statistics are obtained from
two separate GWASs. Specifically, for a group of independent SNPs $Z_{1},\cdots,Z_{p}$, we have the exposure-SNP marginal
regression coefficients and its standard errors (SEs) $\{\hat{\gamma}_j,{\hat{\sigma}}_{{\hat{\gamma}}_{j}}, j=1,\cdots,p\}$ from one GWAS, and the outcome-SNP marginal regression coefficients and its SEs 
$\{\hat{\Gamma}_j,{\hat{\sigma}}_{{\hat{\Gamma}}_{j}}, j=1,\cdots,p\}$ from another GWAS.\cite{Burgess2013}
Following the two-sample summary-data
MR literature,\cite{Zhao2020,Pierce2013,Bowden2017,Ye2021}, the relationships among the SNPs $Z_1,Z_2,\cdots,Z_p$, the exposure $X$, the continuous outcome
$Y$, and the unmeasured confounder $U$, can be formulated by the following linear structure models:
\begin{eqnarray}
	X &=& {\sum\limits_{j = 1}^{p}{\gamma_{j}Z_{j}}} + U + E_{X},  \label{eq1} \\
	Y &=& \beta_{0}X + U + E_{Y}, \label{eq2}
\end{eqnarray}
where $\gamma_{j}$ is the genetic effect of $Z_{j}$ on $X$, $\beta_{0}$ is the causal effect to be estimated, 
and $E_{X}$ and $E_{Y}$ are mutually independent random errors. Let $\Gamma_{j}$ be the genetic effect of 
$Z_{j}$ on $Y$, then we have $\Gamma_{j}=\beta_{0}\gamma_{j}$.
Throughout the paper, we make the following assumptions, which also have been mentioned in other literature:\cite{Zhao2020,Ye2021,Xinwei2023,Xu2023}

\begin{assumption} \label{assum1}
	The sample sizes, $n_X$ and $n_Y$, for the exposure and outcome GWASs, diverge to infinity with the same order.
	The number of independent SNPs, $p$, diverges to infinity. 
\end{assumption}

\begin{assumption} \label{assum2}
	$\{ {{\hat{\gamma}}_{1},\cdots,{\hat{\gamma}}_{p}, {\hat{\Gamma}}_{1},\cdots,{\hat{\Gamma}}_{p}} \}$ are mutually independent. For every $j=1,\cdots,p$, 
	${\hat{\gamma}}_{j} \sim N( {\gamma_{j},\sigma_{{\hat{\gamma}}_{j}}^{2}} )$ 
	and $ {\hat{\Gamma}}_{j} \sim N(\beta_0 \gamma_j,\sigma_{{\hat{\Gamma}}_{j}}^{2})$ 
	with known variances 
	$\sigma_{{\hat{\gamma}}_{j}}^{2}$ and $\sigma_{{\hat{\Gamma}}_{j}}^{2}$, 
	and the variance ratio $\sigma_{{\hat{\gamma}}_{j}}^{2} \big/\sigma_{{\hat{\Gamma}}_{j}}^{2}$ is bounded away from zero and infinity. 
\end{assumption}
Assumptions \ref{assum1} and \ref{assum2} are reasonable since modern GWASs typically enroll large sample sizes, which makes the
normal approximation sufficiently accurate. In addition, the independence among the $2p$ marginal regression coefficients
is guaranteed by the two-sample MR design and the linkage-disequilibrium clumping.\cite{Burgess2013,Purcell2007}

If all SNPs are valid IVs, then each Wald ratio ${\hat{\beta}}_{j} = {{\hat{\Gamma}}_{j}/{\hat{\gamma}}_{j}}$ gives an estimate of the causal effect $\beta_{0}$. 
An intuitive strategy is to aggregate these $p$ ratios, ${\hat{\beta}}_{j}, j = 1,\cdots,p$,
under the meta-analysis framework. 
Many existing methods follow this strategy and the most popular one among them is the inverse-variance weighted (IVW) estimator,\cite{Burgess2013,Bowden2016}
which is given as 
$${\hat{\beta}}_{\rm IVW} = \frac{\sum_{j = 1}^{p}{{\hat{\sigma}}_{{\hat{\Gamma}}_{j}}^{- 2}{\hat{\gamma}}_{j}{\hat{\Gamma}}_{j}}}{\sum_{j = 1}^{p}{{\hat{\sigma}}_{{\hat{\Gamma}}_{j}}^{- 2}{\hat{\gamma}}_{j}^{2}}}.$$
Despite its popularity, studies have pointed out that the IVW estimator can be heavily biased toward zero when there
are many weak IVs.\cite{Zhao2020,Ye2021} Ye et al.\cite{Ye2021} recently proposed the debiased IVW (dIVW) estimator with much better statistical
properties than the IVW estimator, especially in the presence of many weak IVs. The dIVW estimator is written as
$${\hat{\beta}}_{\rm dIVW} = \frac{\sum_{j = 1}^{p}{{\hat{\sigma}}_{{\hat{\Gamma}}_{j}}^{- 2}{\hat{\gamma}}_{j}{\hat{\Gamma}}_{j}}}{\sum_{j = 1}^{p}{{\hat{\sigma}}_{{\hat{\Gamma}}_{j}}^{- 2}\left( {{\hat{\gamma}}_{j}^{2} - {\hat{\sigma}}_{{\hat{\gamma}}_{j}}^{2}} \right)}}.$$
As can be seen, the key idea of the dIVW estimator is to use the unbiased estimator ${\hat{\gamma}}_{j}^{2} - {\hat{\sigma}}_{{\hat{\gamma}}_{j}}^{2}$ of $\gamma_{j}^{2}$
instead of the simple ${\hat{\gamma}}_{j}^{2}$ as the denominator. 
Although ${\hat{\beta}}_{\rm dIVW}$ is proved to be consistent and asymptotically normal under weaker conditions than ${\hat{\beta}}_{\rm IVW}$,it requires the effective sample size greater than 20 to maintain asymptotics.\cite{Ye2021} That is, the bias of  ${\hat{\beta}}_{\rm dIVW}$ may not be
negligible when the effective sample size is small. On the other hand, Zhao et al.\cite{Zhao2020} proposed MR-RAPS from a likelihood
perspective by using a robust adjusted profile score. Although MR-RAPS is also robust to many weak IVs, it does not
have a closed-form solution, and the estimate may even not be unique.

In this paper, we propose a novel modified debiased IVW (mdIVW) estimator by multiplying a modification factor
to the original dIVW estimator. After this simple correction, we prove that the bias of the mdIVW estimator converges to
zero at a faster rate than that of the dIVW estimator under some regularity conditions. Moreover, the mdIVW estimator
has smaller variance than the dIVW estimator, especially in the setting of many weak IVs. Furthermore, the mdIVW
estimator is consistent and asymptotically normal under the same conditions as the dIVW estimator requires. Similarly,
we extend our method to account for the presence of IV selection and balanced horizontal pleiotropy. We investigate
the performance of the proposed mdIVW estimator through extensive simulation studies. Finally, we apply the mdIVW
estimator to estimate the causal effects of coronary artery disease and dyslipidemia on heart failure.

\section{Method}\label{sec2}

\subsection{The modified debiased IVW estimator}
According to Ye et al.,\cite{Ye2021} we define the average IV strength for $p$ IVs as
$$\kappa = \frac{1}{p}{\sum_{j = 1}^{p}{\sigma_{{\hat{\gamma}}_{j}}^{-2}\gamma_{j}^{2}}},$$
which can be estimated by $\hat{\kappa} = \frac{1}{p}{\sum_{j = 1}^{p}{{\hat{\sigma}}_{{\hat{\gamma}}_{j}}^{-2}{\hat{\gamma}}_{j}^{2}}} - 1$. 
We also follow Ye et al.\cite{Ye2021} to define the effective sample size as $\psi=\kappa\sqrt{p}$.
Let ${\hat{\theta}}_{1} = {\sum_{j = 1}^{p}{{\hat{\sigma}}_{{\hat{\Gamma}}_{j}}^{-2}{\hat{\gamma}}_{j}{\hat{\Gamma}}_{j}}}$ 
and ${\hat{\theta}}_{2} = {\sum_{j = 1}^{p}{{\hat{\sigma}}_{{\hat{\Gamma}}_{j}}^{-2}\left( {{\hat{\gamma}}_{j}^{2} - {\hat{\sigma}}_{{\hat{\gamma}}_{j}}^{2}} \right)}}$, 
then the dIVW estimator can be written as a ratio estimator, i.e.,
${\hat{\beta}}_{\rm dIVW} = {{\hat{\theta}}_{1}}/{{\hat{\theta}}_{2}}$.
By Taylor series expansion (see Web Appendix A), we derive the bias of the dIVW estimator as
$${\rm Bias}_{\hat{\beta}_{\rm dIVW}} = \frac{\theta_{1}v_{2}}{\theta_{2}^{3}} - \frac{v_{12}}{\theta_{2}^{2}} + o\left(\varphi\right),$$
where $\theta_{1}=\beta_{0}{\sum_{j = 1}^{p}{\sigma_{{\hat{\Gamma}}_{j}}^{- 2}\gamma_{j}^{2}}}$ is the expectation of $\hat{\theta}_1$,
$\theta_{2}  = {\sum_{j = 1}^{p}{\sigma_{{\hat{\Gamma}}_{j}}^{- 2}\gamma_{j}^{2}}}$ is the expectation of $\hat{\theta}_2$,
$v_{2} = {\sum\limits_{j = 1}^{p}{\sigma_{{\hat{\Gamma}}_{j}}^{- 4}\left( {4\sigma_{{\hat{\gamma}}_{j}}^{2}\gamma_{j}^{2} 
			+ 2\sigma_{{\hat{\gamma}}_{j}}^{4}} \right)}}$ is tha variance of $\hat{\theta}_2$,
 $v_{12} = 2\beta_{0}{\sum\limits_{j = 1}^{p}{\sigma_{{\hat{\Gamma}}_{j}}^{- 4}\sigma_{{\hat{\gamma}}_{j}}^{2}\gamma_{j}^{2}}}$ is the covariance of $\hat{\theta}_1$ and $\hat{\theta}_2$,
and  $\varphi = \left( {\psi\sqrt{p}} \right)^{- 1} + \psi^{- 2}$.
Since the true values of $\theta_1$, $\theta_2$, $v_2$ and $v_{12}$ are unknown, we estimate ${\rm Bias}_{\hat{\beta}_{\rm dIVW}}$ as
$${\widehat{\rm Bias}}_{\hat{\beta}_{\rm dIVW}} = \frac{{\hat{\theta}}_{1}{\hat{v}}_{2}}{{\hat{\theta}}_{2}^{3}} - \frac{{\hat{v}}_{12}}{{\hat{\theta}}_{2}^{2}},$$
where ${\hat{v}}_{2} = {\sum\limits_{j = 1}^{p}{{\hat{\sigma}}_{{\hat{\Gamma}}_{j}}^{- 4}\left( {4{\hat{\sigma}}_{{\hat{\gamma}}_{j}}^{2}{\hat{\gamma}}_{j}^{2} - 2{\hat{\sigma}}_{{\hat{\gamma}}_{j}}^{4}} \right)}}$ 
and ${\hat{v}}_{12} = 2{\sum\limits_{j = 1}^{p}{{\hat{\sigma}}_{{\hat{\Gamma}}_{j}}^{- 4}{\hat{\sigma}}_{{\hat{\gamma}}_{j}}^{2}{\hat{\gamma}}_{j}{\hat{\Gamma}}_{j}}}$ 
are the unbiased estimator of $v_2$ and $v_{12}$, respectively.
Through a first-order bias correction, the proposed mdIVW estimator is defined as
$${\hat{\beta}}_{\rm mdIVW} = {\hat{\beta}}_{\rm dIVW} - {\widehat{\rm Bias}}_{\hat{\beta}_{\rm dIVW}} = \left( {1 - \frac{{\hat{v}}_{2}}{{\hat{\theta}}_{2}^{2}} + \frac{{\hat{v}}_{12}}{{\hat{\theta}}_{1}{\hat{\theta}}_{2}}} \right){\hat{\beta}}_{\rm dIVW},$$
where $1 -{{\hat{v}}_{2}}/{{\hat{\theta}}_{2}^{2}} + {{\hat{v}}_{12}}/({\hat{\theta}}_{1}{\hat{\theta}}_{2})$
can be viewed as a modification factor to the original dIVW estimator. 
For the bias and variance of the mdIVW estimator, we have the following theorem.

\begin{theorem} \label{theom1}
	Suppose Assumptions \ref{assum1} and \ref{assum2} hold, and the effective sample size $\psi\rightarrow\infty$. Then we have the following
	results:
	\begin{enumerate}[(a.)]
		\item  the biases of ${\hat{\beta}}_{\rm dIVW}$ is $O\left(\varphi\right)$, and the bias of ${\hat{\beta}}_{\rm mdIVW}$ is $O\left(\varphi^2\right)$;
		\item The variance difference between ${\hat{\beta}}_{\rm dIVW}$ and ${\hat{\beta}}_{\rm mdIVW}$
			$${\rm Var}\left( {\hat{\beta}}_{\rm dIVW} \right) - {\rm Var}\left( {\hat{\beta}}_{\rm mdIVW} \right) = \frac{2\beta_{0}^{2}\Delta}{\theta_{2}^{4}} + o\left( \varphi^{2} \right),$$
		where
		$$\Delta = \frac{v_{1}v_{2}}{\beta_{0}^{2}} - \frac{6v_{12}v_{2}}{\beta_{0}} + \frac{2v_{12}^{2}}{\beta_{0}^{2}} + 3v_{2}^{2} - 
		\theta_{2}{\sum\limits_{j = 1}^{p}\left\{ {\sigma_{{\hat{\Gamma}}_{j}}^{- 6}\sigma_{{\hat{\gamma}}_{j}}^{6}\left( {6\sigma_{{\hat{\gamma}}_{j}}^{- 2}\gamma_{j}^{2} + 8} \right) + 2\beta_{0}^{- 2}\sigma_{{\hat{\Gamma}}_{j}}^{- 4}\sigma_{{\hat{\gamma}}_{j}}^{4}\left( {\sigma_{{\hat{\gamma}}_{j}}^{- 2}\gamma_{j}^{2} + 1} \right)} \right\}}$$
		and 
		$$v_{1} = {\rm Var}\left( {\hat{\theta}}_{1} \right) = {\sum\limits_{j = 1}^{p}{\sigma_{{\hat{\Gamma}}_{j}}^{- 4}\left( {\beta_{0}^{2}{\sigma_{{\hat{\gamma}}_{j}}^{2}\gamma_{j}^{2}} 
					+ {\sigma_{{\hat{\Gamma}}_{j}}^{2}\gamma_{j}^{2}} + \sigma_{{\hat{\gamma}}_{j}}^{2}\sigma_{{\hat{\Gamma}}_{j}}^{2}} \right)}}.$$
	\end{enumerate}
\end{theorem}

The proof of Theorem \ref{theom1} is given in Appendix A. 
Theorem 1(a) states that the bias of ${\hat{\beta}}_{\rm mdIVW}$ converge to zero 
at a faster rate than that of ${\hat{\beta}}_{\rm dIVW}$
Theorem 1(b) shows that the variance of ${\hat{\beta}}_{\rm mdIVW}$ is also smaller 
than that of ${\hat{\beta}}_{\rm dIVW}$ when $\Delta>0$. 
In fact, we have proved that $\Delta>0$ holds as long as the variance explained by each genetic variant is no more than $1/2$,
which is generally true for complex traits (see more details in Web Appendix B).

According to Ye et al.\cite{Ye2021}, the variance of ${\hat{\beta}}_{\rm dIVW}$ is estimated by 
$$\hat{V}_{\hat{\beta}_{{\rm dIVW}}} = {\hat{\theta}}_{2}^{- 2}{\sum\limits_{j = 1}^{p}\left\{ {{\hat{\sigma}}_{{\hat{\Gamma}}_{j}}^{- 2}{\hat{\gamma}}_{j}^{2} 
		+ {\hat{\beta}}_{\rm dIVW}^{2}{\hat{\sigma}}_{{\hat{\Gamma}}_{j}}^{- 4}\hat{\sigma}_{{\hat{\gamma}}_{j}}^{2}\left( {{\hat{\gamma}}_{j}^{2} 
			+ {\hat{\sigma}}_{{\hat{\gamma}}_{j}}^{2}} \right)} \right\}}$$
Thus, the variance of ${\hat{\beta}}_{\rm mdIVW}$ can be estimated by
	$$\hat{V}_{\hat{\beta}_{{\rm mdIVW}}} = {\hat{\theta}}_{2}^{- 2}{\sum\limits_{j = 1}^{p}\left\{ {{\hat{\sigma}}_{{\hat{\Gamma}}_{j}}^{- 2}{\hat{\gamma}}_{j}^{2} 
		+ {\hat{\beta}}_{\rm mdIVW}^{2}{\hat{\sigma}}_{{\hat{\Gamma}}_{j}}^{- 4}\hat{\sigma}_{{\hat{\gamma}}_{j}}^{2}\left( {{\hat{\gamma}}_{j}^{2} 
			+ {\hat{\sigma}}_{{\hat{\gamma}}_{j}}^{2}} \right)} \right\}} - 2{\hat{\beta}}_{\rm mdIVW}^{2}{\hat{\theta}}_{2}^{- 4}\hat{\Delta},$$
where
$$\hat{\Delta} = \frac{{\hat{v}}_{1}{\hat{v}}_{2}}{{\hat{\beta}}_{\rm mdIVW}^{2}} 
- \frac{{6\hat{v}}_{12}{\hat{v}}_{2}}{{\hat{\beta}}_{\rm mdIVW}} 
+ \frac{2{\hat{v}}_{12}^{2}}{{\hat{\beta}}_{\rm mdIVW}^{2}} 
+ 3{\hat{v}}_{2}^{2} - {\hat{\theta}}_{2}{\sum\limits_{j = 1}^{p}\left\{ {{\hat{\sigma}}_{{\hat{\gamma}}_{j}}^{6}
		{\hat{\sigma}}_{{\hat{\Gamma}}_{j}}^{- 6}\left( {6{\hat{\sigma}}_{{\hat{\gamma}}_{j}}^{- 2}{\hat{\gamma}}_{j}^{2} + 2} \right) 
		+ 2{\hat{\beta}}_{\rm mdIVW}^{- 2}{\hat{\sigma}}_{{\hat{\Gamma}}_{j}}^{- 4}{\hat{\sigma}}_{{\hat{\gamma}}_{j}}^{2}{\hat{\gamma}}_{j}^{2}} \right\}}$$
is the estimate of $\Delta$.
and 
${\hat{v}}_{1} = {\sum\limits_{j = 1}^{p}{{\hat{\sigma}}_{{\hat{\Gamma}}_{j}}^{- 4}( {{\hat{\sigma}}_{{\hat{\gamma}}_{j}}^{2}{\hat{\Gamma}}_{j}^{2} 
			+ {\hat{\sigma}}_{{\hat{\Gamma}}_{j}}^{2}{\hat{\gamma}}_{j}^{2} - {\hat{\sigma}}_{{\hat{\gamma}}_{j}}^{2}{\hat{\sigma}}_{{\hat{\Gamma}}_{j}}^{2}} )}}$ 
is an unbiased estimate of $v_1$.
Obviously, we have $\hat{\beta}_{{\rm mdIVW}}-\hat{\beta}_{{\rm dIVW}}= O_p(\varphi)$.
Therefore, given the asymptotical normality of $\hat{\beta}_{{\rm dIVW}}$, it is straightforward that $\hat{\beta}_{{\rm mdIVW}}$ is also asymptotically normal under the same conditions required by  $\hat{\beta}_{{\rm dIVW}}$.

Note that the validity of the above theorems relies on the condition that the effective sample size $\psi\rightarrow\infty$. 
As can be seen, even if there are many weak IVs, in which case $\kappa\rightarrow0$, Theorem \ref{theom1} still holds as long as the number of such weak IVs $p$ is large enough.

\subsection{Selection of Candidate Instruments}
In this section, we explore the performance of the mdIVW estimator with IV selection. 
As Zhao et al.\cite{Zhao2020} pointed out,
using extremely weak or even null IVs may increase the variance of the MR estimator and thus make it less efficient. When
summary statistics from another independent exposure GWAS are available, it is a common practice to screen out weak
IVs. We use $\left\{ {{\hat{\gamma}}_{j}^{\ast},\sigma_{{\hat{\gamma}}_{j}}^{\ast}}, j =1,\cdots,p\right\}$ 
to represent the summary statistics obtained from the selection dataset. 
Similarly, we have the following assumption for the above summary statistics.

\begin{assumption} \label{assum3}
	$\left\{ {{\hat{\gamma}}_{1},\cdots,{\hat{\gamma}}_{p},~{\hat{\Gamma}}_{1},\cdots,{\hat{\Gamma}}_{p},{\hat{\gamma}}_{1}^{\ast},\cdots,{\hat{\gamma}}_{p}^{\ast}} \right\}$ 
	are mutually independent.
	For every $j = 1,\cdots,p$, $\left. {\hat{\gamma}}_{j}^{\ast} \right.\sim N\left( {\gamma_{j},\sigma_{{\hat{\gamma}}_{j}}^{\ast2}} \right)$ with known variance 
	$\sigma_{{\hat{\gamma}}_{j}^\ast}^{2}$ and the variance ratio $\sigma_{{\hat{\gamma}}_{j}}^{2}/\sigma_{{\hat{\gamma}}_{j}^\ast}^{2}$ is bounded away from zero and infinity.
\end{assumption}

In practice, only the IVs with $\left |\hat{\gamma}_{j} ^{\ast}   \right | / \sigma_{\hat{\gamma}_{j}^\ast} > \lambda $ are included in the subsequent MR analysis, 
where $\lambda$ is a predetermined threshold. 
Researchers typically use 5.45 as the value of $\lambda$, which corresponds to the genome-wide significance level of $5\times{10}^{-8}$. 
However, Ye et al.\cite{Ye2021} show that this threshold is extremely stringent such that some valid IVs may be excluded from the analysis, 
thereby reducing efficiency. They also recommend a threshold $\lambda=\sqrt{2{\rm log} p}$ to guarantee a small probability of selecting any invalid IVs.
Following Ye et al.,\cite{Ye2021} we define the average IV strength with screening threshold $\lambda$ as
$$\kappa_{\lambda} = \frac{1}{p_{\lambda}}{\sum_{j = 1}^{p}{\sigma_{{\hat{\gamma}}_{j}}^{- 2}\gamma_{j}^{2}q_{\lambda,j}}},$$
where $q_{\lambda,j} = P( {{\left| {\hat{\gamma}}_{j}^{\ast} \right|/\sigma_{{\hat{\gamma}}_{j}}^{\ast}} > \lambda} )$ 
and $p_{\lambda} = {\sum_{j = 1}^{p}q_{\lambda,j}}$.
Let $s_{\lambda,j} =I ( \left | \hat{\gamma}_j^\ast \right | / \sigma _{{\hat{\gamma} _j}^ \ast}  > \lambda ) $
where $I\left ( \cdot  \right ) $ is the indicator function.
Then $\kappa_\lambda$ can be estimated as
$$\hat{\kappa}_\lambda=\frac{1}{{\hat{p}}_\lambda}\sum_{j=1}^{p}{{\hat{\sigma}}_{{\hat{\gamma}}_j}^{-2}{\hat{\gamma}}_j^2s_{\lambda,j}}-1,$$
where ${\hat{p}}_\lambda=\sum_{j=1}^{p}s_{\lambda,j}$.
In the case of IV selection, we redefine the effective sample size
$\psi_\lambda=\kappa_\lambda\sqrt{p_\lambda}/{\rm max}(1,\omega)$ and $\varphi_\lambda = \left(\psi_\lambda\sqrt{p_\lambda}{\rm max}(1,\omega) \right)^{-1} + \psi_\lambda^{-2}$,
where $\omega^{2} =\big(  {p_{\lambda}^{- 1}}{\sum\limits_{j = 1}^{p}{\sigma_{{\hat{\Gamma}}_{j}}^{- 4}\gamma_{j}^{4}q_{\lambda,j}\left( {1 - q_{\lambda,j}} \right)}}\big) ^ {-1/2}$.
Formally, the mdIVW estimator with IV selection is expressed as
$${\hat{\beta}}_{\lambda,{\rm mdIVW}} = \left( {1 - \frac{{\hat{v}}_{2,\lambda}}{{\hat{\theta}}_{2,\lambda}^{2}} + \frac{{\hat{v}}_{12,\lambda}}{{\hat{\theta}}_{1,\lambda}{\hat{\theta}}_{2,\lambda}}} \right)\frac{{\hat{\theta}}_{1,\lambda}}{{\hat{\theta}}_{2,\lambda}} = \left( {1 - \frac{{\hat{v}}_{2,\lambda}}{{\hat{\theta}}_{2,\lambda}^{2}} + \frac{{\hat{v}}_{12,\lambda}}{{\hat{\theta}}_{1,\lambda}{\hat{\theta}}_{2,\lambda}}} \right){\hat{\beta}}_{\lambda,{\rm dIVW}},$$
where ${\hat{\theta}}_{1,\lambda} = {\sum\limits_{j = 1}^{p}{{\hat{\sigma}}_{{\hat{\Gamma}}_{j}}^{- 2}{\hat{\gamma}}_{j}{\hat{\Gamma}}_{j}}}s_{\lambda,j}$ 
and ${\hat{\theta}}_{2,\lambda} = {\sum\limits_{j = 1}^{p}{{\hat{\sigma}}_{{\hat{\Gamma}}_{j}}^{- 2}\left( {{\hat{\gamma}}_{j}^{2} - {\hat{\sigma}}_{{\hat{\gamma}}_{j}}^{2}} \right)s_{\lambda,j}}}$
are the numerator and denominator of the dIVW estimator with IV selection, respectively, 
${\hat{v}}_{12,{\lambda}} = 2{\sum\limits_{j = 1}^{p}{{\hat{\sigma}}_{{\hat{\Gamma}}_{j}}^{- 4}{\hat{\sigma}}_{{\hat{\gamma}}_{j}}^{2}}}{{\hat{\gamma}}_{j}\hat{\Gamma}}_{j}s_{\lambda,j}$
is the extimate of 
${\rm Cov}( {{\hat{\theta}}_{1,\lambda},{\hat{\theta}}_{2,\lambda}} )$,
and $\hat{v}_{2,\lambda} = {\sum\limits_{j = 1}^{p}{{\hat{\sigma}}_{{\hat{\Gamma}}_{j}}^{- 4}( {4{\hat{\sigma}}_{{\hat{\gamma}}_{j}}^{2}{\hat{\gamma}}_{j}^{2} - 2{\hat{\sigma}}_{{\hat{\gamma}}_{j}}^{4}} )s_{\lambda,j}}}$
is the estimate of 
${\rm Var}({\hat{\theta}}_{2,\lambda})$.
The following theorem shows that the mdIVW estimator still has
smaller bias and variance than the dIVW estimator in the presence of IV selection.

\begin{theorem} \label{theom2}
	Suppose the assumptions \ref{assum1}-\ref{assum3} hold 
	and that the effective sample size $\psi_\lambda\rightarrow\infty$. Then we have th following results:
	\begin{enumerate}[(a.)]
		\item The biases of ${\hat{\beta}}_{\lambda,{\rm dIVW}}$ is  $O\left(\varphi_\lambda\right)$, and the bias of  ${\hat{\beta}}_{\lambda,{\rm mdIVW}}$  is $O\left({\varphi_\lambda^2}\right)$;
		\item The variance difference between  ${\hat{\beta}}_{\lambda,{\rm dIVW}}$  and  ${\hat{\beta}}_{\lambda,{\rm mdIVW}}$  is
		$${\rm Var}\left( {\hat{\beta}}_{\lambda,{\rm dIVW}} \right) - {\rm Var}\left( {\hat{\beta}}_{\lambda,{\rm mdIVW}} \right) 
		= \frac{2\beta_{0}^{2}}{\theta_{2,\lambda}^{4}}\Delta_{\lambda} + O\left( \varphi_{\lambda}^{3} \right),$$
		
	\end{enumerate}
\end{theorem}
The proof of Theorem \ref{theom2} and the detailed expression of $\Delta_\lambda$ are both given in Web Appendix C. We also prove that $\Delta_\lambda > 0 $ holds in a similar manner (see details in Web Appendix D), so that the variance of ${\hat{\beta}}_{\lambda,{\rm mdIVW}}$ is smaller than that of ${\hat{\beta}}_{\lambda,{\rm dIVW}}$.

Similar to the estimate of the variance of ${\hat{\beta}}_{\lambda,{\rm dIVW}}$, the variance of ${\hat{\beta}}_{\lambda,{\rm mdIVW}}$ can be estimated by
$${\hat{V}}_{{\hat{\beta}}_{\lambda,{\rm mdIVW}}} = {\hat{\theta}}_{2,\lambda}^{- 2}{\sum\limits_{j = 1}^{p}{\left\{ {{\hat{\sigma}}_{{\hat{\Gamma}}_{j}}^{- 2}{\hat{\gamma}}_{j}^{2} + {\hat{\beta}}_{\lambda,{\rm mdIVW}}^{2}{\hat{\sigma}}_{{\hat{\Gamma}}_{j}}^{- 4}{\hat{\sigma}}_{{\hat{\gamma}}_{j}}^{2}\left( {{\hat{\gamma}}_{j}^{2} + {\hat{\sigma}}_{{\hat{\gamma}}_{j}}^{2}} \right)} \right\} s_{\lambda,j}}} - 2{\hat{\beta}}_{\lambda,{\rm mdIVW}}^{2}{\hat{\theta}}_{2,\lambda}^{- 4}{\hat{\Delta}}_{\lambda},$$
where 
$${\hat{\Delta}}_{\lambda} =  {\frac{{\hat{v}}_{1,\lambda}{\hat{v}}_{2,\lambda}}{{\hat{\beta}}_{\lambda,{\rm mdIVW}}^{2}} 
	- \frac{6{\hat{v}}_{12,\lambda}{\hat{v}}_{2,\lambda}}{{\hat{\beta}}_{\lambda,{\rm mdIVW}}} 
	+ \frac{2{\hat{v}}_{12,\lambda}^{2}}{{\hat{\beta}}_{\lambda,{\rm mdIVW}}^{2}} 
	+ 3{\hat{v}}_{2,\lambda}^{2} - 
	{\hat{\theta}}_{2,\lambda}{\sum\limits_{j = 1}^{p}{\left\{ {\hat{\sigma}}_{{\hat{\Gamma}}_{j}}^{- 6}
			{{\hat{\sigma}}_{{\hat{\gamma}}_{j}}^{6}\left( {6{\hat{\sigma}}_{{\hat{\gamma}}_{j}}^{- 2}{\hat{\gamma}}_{j}^{2} + 2} \right)} \right\}s_{\lambda,j}}} 
	- {2{\hat{\theta}}_{2,\lambda}}{{\hat{\beta}}_{\lambda,{\rm mdIVW}}^{-2}}{\sum\limits_{j = 1}^{p} {{\hat{\sigma}}_{{\hat{\Gamma}}_{j}}^{- 4}
			{\hat{\sigma}}_{{\hat{\gamma}}_{j}}^{2}{\hat{\gamma}}_{j}^{2}s_{\lambda,j}} }}$$
	is the estimate of $\Delta_\lambda$, and
 ${\hat{v}}_{1,\lambda} = {\sum\limits_{j = 1}^{p}{{\hat{\sigma}}_{{\hat{\Gamma}}_{j}}^{- 4}( {{\hat{\sigma}}_{{\hat{\gamma}}_{j}}^{2}{\hat{\Gamma}}_{j}^{2} 
			+ {\hat{\sigma}}_{{\hat{\Gamma}}_{j}}^{2}{\hat{\gamma}}_{j}^{2} - {\hat{\sigma}}_{{\hat{\gamma}}_{j}}^{2}{\hat{\sigma}}_{{\hat{\Gamma}}_{j}}^{2}}
		)s_{\lambda,j}}}$ is the estimate of ${\rm Var}({\hat{\theta}}_{1,\lambda})$.
Similar to the asymptotical normality of ${\hat{\beta}}_{\rm mdIVW}$, we have 
${\hat{\beta}}_{\lambda,{\rm mdIVW}} - {\hat{\beta}}_{\lambda,{\rm dIVW}} = O_p(\varphi_\lambda)$.
Thus, ${\hat{\beta}}_{\lambda,{\rm mdIVW}}$ requires no more conditions than  ${\hat{\beta}}_{\lambda,{\rm dIVW}}$ to
maintain asymptotical normality.
In fact, when $\lambda=0$ we have $q_{\lambda,j}=1$, that is ${\hat{\beta}}_{\rm mdIVW}$ is a special case of ${\hat{\beta}}_{\lambda,{\rm mdIVW}}$ with $\lambda=0$.

\subsection{Accounting for balanced horizontal pleiotropy}
We extend the mdIVW estimator to a common pleiotropy setting, known as the balanced horizontal pleiotropy. 
In this scenario, we need to modify the linear structural model as
\begin{align}
Y = \beta_{0}X + {\sum_{j = 1}^{p}{\alpha_{j}Z_{j}}} + U + E_{Y}, \label{eq3}
\end{align}
where $\alpha_j\sim N(0,\tau_0^2)$ is the pleiotropic effect of $Z_j$, which is assumed to be independent of $\gamma_j$ and $E_Y$.
In this scenario, we have $\Gamma_j=\beta_0\gamma_j+\alpha_j$.
When treating $\alpha_j$ as a random effect, we have ${\hat{\Gamma}}_j \sim N(\beta_0\gamma_j,\sigma_{{\hat{\Gamma}}_j}^2+\tau_0^2)$. Thus, we rewrite Assumption \ref{assum2} to account for the balanced horizontal pleiotropy.
\begin{assumption} \label{assum4}
	Assumption 2 holds except $ {\hat{\Gamma}}_{j} \sim N(\beta_0\gamma_j,\sigma_{{\hat{\Gamma}}_{j}}^{2} + \tau_{0}^{2})$,
	for $j=1,\cdots,p$. In addition,	for some constant $c_+$, $\tau_0<c_+\sigma_{{\hat{\Gamma}}_j}$ for all $j$. 
\end{assumption}

According to Ye et al.,\cite{Ye2021} $\tau_0^2$ can be estimated as follows:\cite{Ye2021}
$${\hat{\tau}}^{2} = \frac{\sum\limits_{j = 1}^{p}{\Big\{ {\big( {{\hat{\Gamma}}_{j} - {\hat{\beta}}_{\rm mdIVW}\hat{\gamma}_j} \big)^{2} - {{\sigma}}_{{\hat{\Gamma}}_{j}}^{2} - {\hat{\beta}}_{\rm mdIVW}^{2}{{\sigma}}_{{\hat{\gamma}}_{j}}^{2}} \Big\}{{\sigma}}_{{\hat{\Gamma}}_{j}}^{- 2}}}{\sum_{j = 1}^{p}{{\sigma}}_{{\hat{\Gamma}}_{j}}^{- 2}}.$$
When the balanced horizontal pleiotropy exists, we need to modify the estimate of ${\rm Var}(\hat{\theta}_{1,\lambda})$ as
$${\hat{v}}_{1,\lambda \tau} = {\sum\limits_{j = 1}^{p}{{{\sigma}}_{{\hat{\Gamma}}_{j}}^{- 4}\left\{ {{\sigma}}_{{\hat{\gamma}}_{j}}^{2}{{\hat{\Gamma}}_{j}^{2} + \big({{{\sigma}}_{{\hat{\Gamma}}_{j}}^{2} + {\hat{{\tau}}}^{2}}\big){\hat{\gamma}}_{j}^{2} - \big({{{\sigma}}_{{\hat{\Gamma}}_{j}}^{2} + {\hat{{\tau}}}^{2}} \big){{\sigma}}_{{\hat{\gamma}}_{j}}^{2}}\right\}s_{\lambda,j}}}.$$
Then, the estimate of ${\rm Var}({\hat{\beta}}_{\lambda,{\rm mdIVW}})$ under the balanced pleiotropy is written as
 $${\widehat{V}}_{{\hat{\beta}}_{\lambda,{\rm mdIVW}},\tau} = {\hat{\theta}}_{2,\lambda}^{- 2}{\sum\limits_{j = 1}^{p}{\left\{ {{{\sigma}}_{{\hat{\Gamma}}_{j}}^{- 2}{\hat{\gamma}}_{j}^{2}\big({1 + {\sigma}_{\hat{\Gamma}_{j}}^{- 2}} {\hat{\tau}}^{2}\big) + {\hat{\beta}}_{\lambda,{\rm mdIVW}}^{2}{{\sigma}}_{{\hat{\Gamma}}_{j}}^{- 4}{{\sigma}}_{{\hat{\gamma}}_{j}}^{2}\big( {{\hat{\gamma}}_{j}^{2} + {{\sigma}}_{{\hat{\gamma}}_{j}}^{2}} \big)} \right\} s_{\lambda,j}}} - 2{\hat{\beta}}_{\lambda,{\rm mdIVW}}^{2}{\hat{\theta}}_{2,\lambda}^{- 4}{\widehat{\lambda}}_{\lambda,\tau},$$
where 
$${\widehat{\lambda}}_{\lambda,\tau} = \frac{{\hat{v}}_{1,\lambda\tau}{\hat{v}}_{2,\lambda}}{{\hat{\beta}}_{\lambda,{\rm mdIVW}}^{2}} 
- \frac{6{\hat{v}}_{2,\lambda}{\hat{v}}_{12,\lambda}}{{\hat{\beta}}_{\lambda,{\rm mdIVW}}} 
+ \frac{2{\hat{v}}_{12,\lambda}^{2}}{{\hat{\beta}}_{\lambda,{\rm mdIVW}}^{2}} 
+ 3{\hat{v}}_{2,\lambda}^{2} 
- {\hat{\theta}}_{2,\lambda}{\sum\limits_{j = 1}^{p}{\left\{ {{{\sigma}}_{{\hat{\Gamma}}_{j}}^{- 6}{{\sigma}}_{{\hat{\gamma}}_{j}}^{4}\big({6{\hat{\gamma}}_{j}^{2} + 8{{\sigma}}_{{\hat{\gamma}}_{j}}^{2}} \big) + 2{\hat{\beta}}_{\lambda,{\rm mdIVW}}^{- 2} {{{\sigma}}_{{\hat{\Gamma}}_{j}}^{- 4}{{\sigma}}_{{\hat{\gamma}}_{j}}^{2} \big( 1 + {{\sigma}}_{{\hat{\Gamma}}_{j}}^{-2}{\hat{{\tau}}}^{2}} \big)\big({\hat{\gamma}}_{j}^{2}+\sigma_{\hat{\gamma}_j}^2\big)} \right\} s_{\lambda,j}}}$$ 
is an estimate of $\Delta_\lambda$ under the balanced horizontal pleiotropy. 
In fact, when there is no horizontal pleiotropy $(\tau_0^2=0)$, we have ${\widehat{V}}_{{\hat{\beta}}_{\lambda,{\rm mdIVW}},\tau} = {\widehat{V}}_{{\hat{\beta}}_{\lambda,{\rm mdIVW}}}$. That is, the absence of horizontal pleiotropy can be considered as a special case of the balanced horizontal pleiotropy with $\tau_0^2=0$. Under the above assumptions, Theorems \ref{theom1}-\ref{theom2} can be extended to the case of balanced horizontal pleiotropy (see \textcolor{black}{Web Appendices A and C}).

\section{Simulation studies} \label{sec3}
\subsection{Simulation settings} \label{sec3.1}
We have already analytically proved that both the bias and variance of the mdIVW estimator are simultaneously smaller than those of the dIVW estimator. 
To validate these findings, we further perform extensive simulations by considering various settings.
We generate the summary data for 1000 IVs from 
$ {\hat{\gamma}}_{j} \sim N( {\gamma_{j},\sigma_{{\hat{\gamma}}_{j}}^{2}})$ 
and 
${\hat{\Gamma}}_{j} \sim N( {\beta_{0}\gamma_{j} + \alpha_{j},\sigma_{{\hat{\Gamma}}_{j}}^{2}} )$, 
where $ \alpha_{j} \sim N( 0,\tau_{0}^{2} )$ represents pleiotropy.
Following the simulation settings in Ye et al.,\cite{Ye2021}
we consider the situations with many weak IVs and many null IVs.\cite{Ye2021}
For the first $s$ IVs, 
we draw $\gamma_1,\cdots,\gamma_s$ independently from $N\left( {0,~\sigma^{2}} \right)$.
For the rest IVs, we set $\gamma_j=0, j=s+1,\cdots,1000$. 
We fix the true causal effect $\beta_0$ at 0.5. 
Let $\tau_0$ to be 0 and 0.01, corresponding to the
scenarios of no pleiotropy and balanced horizontal pleiotropy, respectively.
The variances $\sigma_{{\hat{\gamma}}_j}^2$ and $\sigma_{\hat{\Gamma}_j}^2$ are calculated as follows:
$$\sigma_{{\hat{\gamma}}_j}^2=\left({\rm Var} (X)-\gamma_j^2{\rm Var} (Z_j )\right)\big/\left(n_X{\rm Var} (Z_j )\right),$$
$$\sigma_{\hat{\Gamma}_j}^2=\left({\rm Var}(Y)-\beta_0^2\gamma_j^2{\rm Var}(Z_j)\right)\big/\left(n_Y{\rm Var} (Z_j )\right).$$
Let $Z_j$ follow a binomial distribution ${\rm Bin} (2, {\rm MAF}_j )$. We randomly generate ${\rm MAF}_j$, the minor allele frequency of $Z_j$, from the uniform distribution $U (0.1,0.5 )$. Thus, we have ${\rm Var} (Z_j )=2{\rm MAF}_j (1-{\rm MAF}_j )$. 
We set the variances of $U$, $E_X$ and $E_Y$ all to 2. Then, ${\rm Var} (X )$ and ${\rm Var} (Y )$ can be evaluated using Equations  (\ref{eq1}) (see Section 2) and (\ref{eq3}) (see Section 4.3).
Given the parameters $(\gamma_1,\cdots,\gamma_{1000},\sigma_{{\hat{\gamma}}_1}^2,\cdots,\sigma_{{\hat{\gamma}}_{1000}}^2)$
and  $(\Gamma_1,\cdots,\Gamma_{1000},\sigma_{{\hat{\Gamma}}_1}^2,\cdots,\sigma_{{\hat{\Gamma}}_{1000}}^2)$, we can randomly draw summary-data
	$({\hat{\gamma}}_1,\cdots,{\hat{\gamma}}_{1000})$ and $({\hat{\Gamma}}_1,\cdots,{\hat{\Gamma}}_{1000})$ from the normal distributions $N(\gamma_j,\sigma_{{\hat{\gamma}}_j}^2)$ and $ (\Gamma_j,\sigma_{{\hat{\Gamma}}_j}^2 )$, respectively. 
In addition, in scenarios involving IV selection, we set the sample size of the selection dataset, $n_X^\ast$, to be half of $n_X$. Thus, we have ${\hat{\gamma}}_j^\ast \sim  N(\gamma_j,2\sigma_{{\hat{\gamma}}_j}^2)$.
 We fix the IV selection threshold at $\delta=\sqrt{2{\rm log}p}$, which is 3.72 in our simulation settings.

\subsection{A full comparison among the dIVW, pIVW, and mdIVW estimators}
We have proved that the mdIVW estimator has some theoretical advantages. To validate these findings,  we perform extensive simulations by considering the following settings:
\begin{enumerate}[1.]
	\item $s$ varies from 0 to 1000 in the increment of 10;
	\item $\sigma^2$ varies from $1\times10^{-4}$ to $1\times10^{-3}$ in the increment of $1\times10^{-4}$;
	\item $n_X=2n_Y$ varies from 100,000 to 200,000 in the increment of 10,000.
\end{enumerate}
We examine a total of 11,110 parameter combinations. 
For each combination, we repeat 10,000 times to evaluate  the
relative biases (bias divided by $\beta_0$),  
SEs,
MSEs, 
and coverage probabilities (CPs) of the 95\% confidence interval (CI) of  the dIVW, and mdIVW estimators.

\begin{figure}[htb]
	\centering
	\includegraphics[width=1\textwidth]{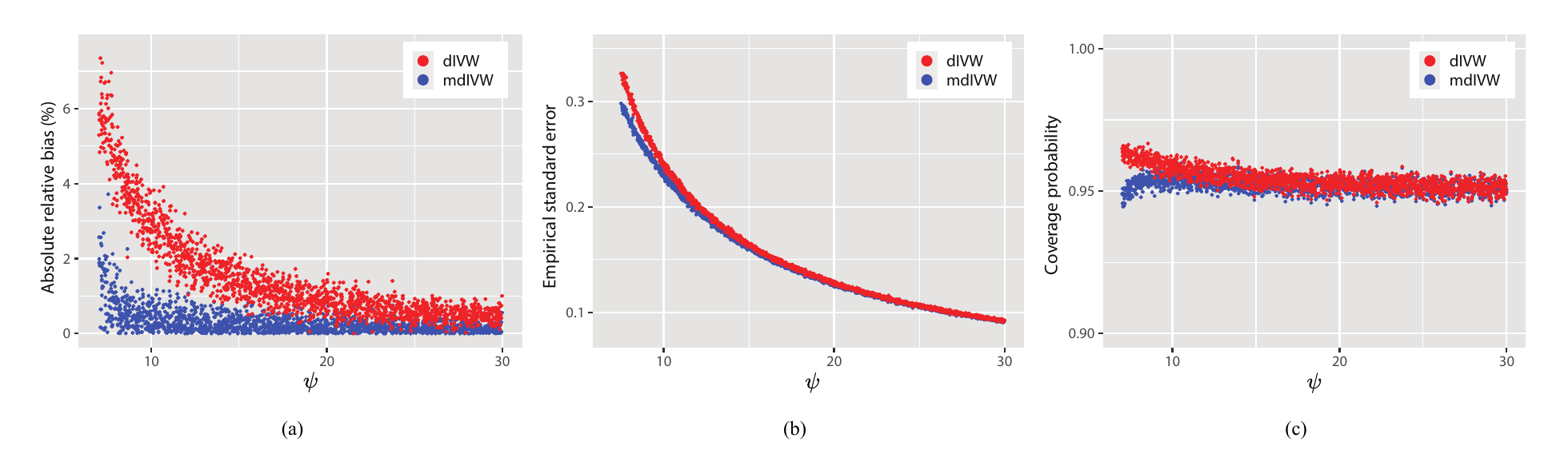}
	\caption{The plots of (a) the estimated relative biases (biases divided by $\beta_0$); 
		(b) the empirical standard errors; 
		and (c) the estimated coverage probabilities of the 95\% confidence intervals for the dIVW and mdIVW estimators versus the mean of empirical effective sample sizes. 
		Each point represents the simulation results of a unique parameter combination based on 10,000 repetitions.
		There is no IV selection or horizontal pleiotropy ($\tau_0=0$).}
	\label{fig2}
\end{figure}

\begin{figure}[htb]
	\centering
	\includegraphics[width=1\textwidth]{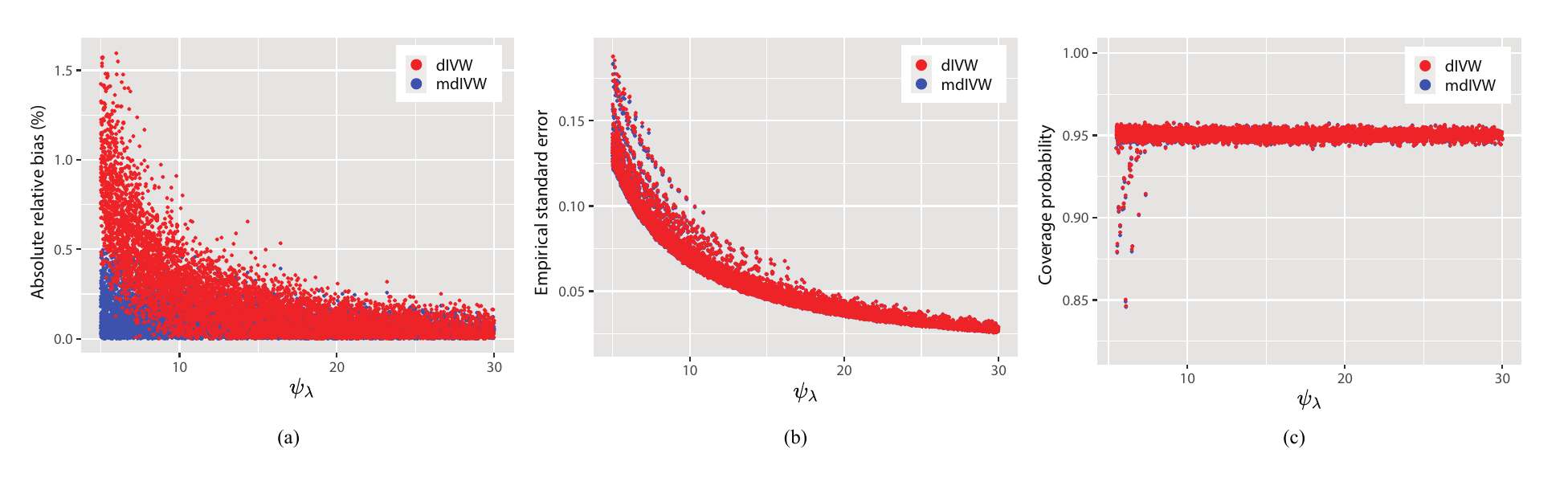}
	\caption{
			The plots of (a) the estimated relative biases (biases divided by $\beta_0$); 
			(b) the empirical standard errors; 
			and (c) the estimated coverage probabilities of the 95\% confidence intervals for the dIVW and mdIVW estimators versus the mean of empirical effective sample sizes. 
			Each point represents the simulation results of a unique parameter combination based on 10,000 repetitions.
			There is no horizontal pleiotropy $(\tau_0=0)$. The IV selection threshold $\delta=\sqrt{2{\rm log}p}$.}
	\label{fig3}
\end{figure}

Figure \ref{fig2} shows the results  of no IV selection and no horizontal pleiotropy. 
From Figure 2a, we can see that the relative bias of ${\hat{\beta}}_{\rm mdIVW}$ is generally smaller than that of ${\hat{\beta}}_{dIVW}$ at a fixed $\psi$. 
The relative bias of ${\hat{\beta}}_{mdIVW}$ is less than 1\% when $\psi > 10$,
but ${\hat{\beta}}_{dIVW}$ requires $\psi>20$. 
Figure 2b shows that the SE of ${\hat{\beta}}_{\rm mdIVW}$ is slightly smaller than that of ${\hat{\beta}}_{\rm dIVW}$. 
The confidence interval derived from the normal approximation of ${\hat{\beta}}_{\rm mdIVW}$
can maintain the nominal coverage probability when $\psi>10$ (see Figure 2c).
Figure \ref{fig3} plots the results with the IV selection but in the absence of horizontal pleiotropy. 
In this case
${\hat{\beta}}_{\rm mdIVW}$ still have smaller bias than ${\hat{\beta}}_{\rm dIVW}$. 
However, both the SEs of ${\hat{\beta}}_{\rm mdIVW}$ and ${\hat{\beta}}_{\rm dIVW}$ are very small , and thus hard to distinguish from each other. 
and thus it is
difficult to distinguish their differences. Compared with Figure 2, we can see that IV selection improves the performances
of the mdIVW and dIVW estimators with reduced bias and variance. The results in the presence of balanced horizontal
pleiotropy (see Web Figures 1-2) show the similar patterns as those in Figure 2 and Figure 3, except with larger variance.

\subsection{Comparison with other commonly used methods}
In this subsection, we additionally compare the mdIVW estimator to several commonly used methods, including the IVW\cite{Burgess2013}, the MR-Median,\cite{Bowden2016} the MR-Egger,\cite{Bowden2015} and the MR-RAPS.\cite{Zhao2020} We fix $n_X=2n_Y=150,000$,
and $\sigma^2=5\times{10}^{-4}$.  
In the case of no IV selection, we set the number of weak IVs $s$ to 50, 100 and 150.  In the case with IV selection, we fix $s$ to 150 and set the sample size of the selection GWAS $n_X^\ast$ to 75,000, 100,000 and 150,000. 
The other parameters are set to be the same as those in subsection \ref{sec3.1}. We repeat 20,000 times for each parameter combination. 

Table \ref{tab1} presents the results in the case of no IV selection and no horizontal pleiotropy. From this table, we find
that the mdIVW estimator consistently has the smallest bias in all three settings. In contrast, the IVW, MR-Median, and
MR-Egger estimators can be seriously biased toward zero and also have very poor coverage probabilities.
The MR-RAPS
estimator also has large bias and SE when the effective sample size $\psi$ is very small ($\psi=8.13$).
Nevertheless, as the effective sample size  $\psi$ increases, the MR-RAPS estimator, dIVW estimator, and mdIVW estimator tend to have similar
performances. As expected, the mdIVW estimator has smaller SE compared to the dIVW estimator in all three settings.
Table \ref{tab2} provides the results with IV selection at threshold $\lambda = \sqrt{2{\rm log}p} = 3.72$ but in the absence of horizontal pleiotropy.
It is observed that the mdIVW estimator still has overall best performance in terms of bias and SE. Similar results can
also be shown when there exists balanced horizontal pleiotropy (see Web Tables 1-2).

\section{Real data analysis} \label{sec4}
Heart failure (HF) is a complex and life-threatening disease that affects more than 30 million people worldwide and thus poses a significant public health challenge.\cite{Ziaeian2016} Several cardiovascular and systemic disorders, such as coronary artery disease (CAD), obesity, and diabetes, are known as etiological factors.\cite{Paul2022} In this section, we aim to estimate the causal effects of CAD and dyslipidemia on HF. The selection and exposure datasets are respectively obtained from two different GWAS meta-analyses: the Genetic Epidemiology Research on Adult Health and Aging (GERA) with 53,991 individuals and the subgroup of UK BioBank with 108,039 individuals.\cite{Zhu2018} The outcome dataset is obtained from the Heart Failure Molecular Epidemiology for Therapeutic Targets (HERMES) consortium, which includes up to 47,309 cases and 930,014 controls from 29 distinct studies.\cite{Shah2020} A more detailed description of the data is given in Web Table 3. We employ the software "PLINK" to obtain linkage-disequilibrium (LD) independent SNPs ($r^2 < 0.001$ within 10 Mb pairs). Finally, we identify 2413 eligible SNPs for CAD and 2480 eligible SNPs for dyslipidemia (\textcolor{black}{see Web Table 4} for details).

We apply the proposed mdIVW estimator along with some commonly used MR methods to analyze the data, and the results are given in Table \ref{tab3}. 
All the methods showed positive causal effects of CAD and dyslipidemia on HF at the significance level of 0.05 
except for the MR-Median estimate of CAD on HF. 
Our findings are in line with the academic consensus.\cite{Paul2022} 
For CAD with a very small $\hat{\psi}$, the MR-RAPS, dIVW, and mdIVW estimators all gave similar estimates before and after IV selection. We also observe that the mdIVW estimator consistently exhibit smaller SE compared to the dIVW estimator, with or without IV selection. 
In contrast, the IVW, MR-Median, and MR-Egger estimators without IV selection are quite different from those with IV selection, indicating that these methods are sensitive to weak IVs.
For dyslipidemia with a relatively large $\hat{\psi}$, the mdIVW, and dIVW estimators yield nearly identical results,  which is consistent with our simulation studies. Even so, the IVW, MR-Median, and MR-Egger estimators are still very close to zero under the case without IV selection.
We also conduct the analysis by assuming the presence of balanced horizontal pleiotropy. The results are given in \textcolor{black}{Web Table 5}, which are similar to those in Table 3. Following Zhao et al.,\cite{Zhao2020} we use the Quantile-Quantile plot of the standardized residuals,
 $$\frac{{\hat{{\Gamma}}}_j-{\hat{\beta}}_{\rm mdIVW}{\hat{\gamma}}_j}{\sqrt{{{\sigma}}_{{\hat{{\Gamma}}}_j}^2+{\hat{\beta}}_{\rm mdIVW}^2{{\sigma}}_{{\hat{\gamma}}_j}^2}},$$
to evaluate the plausibility of Assumption \ref{assum2}, as depicted in \textcolor{black}{Figure 4}. Since the proximity of the residuals to the diagonal, Assumption 2 is likely to hold in this example. 

\section{Discussion} \label{sec5}
Weak IV bias is one of the major challenges in MR analysis. In this article, we have shown that a simple modification
to the dIVW estimator can further reduce the bias and variance simultaneously.We further evaluate the robustness of such
findings by considering the scenarios where the IV selection is conducted or the balanced horizontal pleiotropy presents.
Simulation studies and real data analysis validate the improvement of the proposed mdIVW estimator in terms of bias and
SE. The proposed mdIVW estimator is asymptotically normal even in the presence of many weak IVs, requiring no more
assumptions than the dIVW estimator. Moreover, the mdIVW estimator has a simple closed form and is computationally
simple as well. In contrast, although the MR-RAPS estimator is also robust to weak IVs, it does not have a closed-form
solution and might have multiple solutions. Hence, we recommend using the mdIVW estimator without IV selcetion
as another baseline estimator in two-sample summary-data MR studies. Simulation results suggest that the mdIVW can
maintain good asymptotics as long as the effective sample size is greater than 10.

\bmsection*{Data availability statement}
Summary data that support the findings of this study are available at the following links:
the selection data and exposure data \href{https://cnsgenomics.com/content/data}{https://cnsgenomics.com/content/data};
and the outcome data \href{https://cvd.hugeamp.org/}{https://cvd.hugeamp.org/}.

\bmsection*{Conflict of interest}

The authors declare no potential conflict of interests.

\bmsection*{Acknowledgment}
The authors would like to thank the editor, the associate editor, and the anonymous referees for their constructive comments which greatly improved this article.

\bibliography{wileyNJD-AMA}

\begin{thebibliography}{10}
\providecommand \doibase [0]{http://dx.doi.org/}%

\bibitem{Zhao2020}
Zhao Q, Wang J, Hemani G, Bowden J, Small DS. {Statistical inference in
  two-sample summary-data Mendelian randomization using robust adjusted profile
  score}. {\it The Annals of Statistics.} 2020\string;48(3)\string:1742 --
  1769.
\newblock \href {\doibase 10.1214/19-AOS1866} {doi: 10.1214/19-AOS1866}

\bibitem{San2007}
Sanderson S, Tatt ID, Higgins JP. {Tools for assessing quality and
  susceptibility to bias in observational studies in epidemiology: a systematic
  review and annotated bibliography}. {\it International Journal of
  Epidemiology.} 2007\string;36(3)\string:666-676.
\newblock \href {\doibase 10.1093/ije/dym018} {doi: 10.1093/ije/dym018}

\bibitem{Burgess2017}
Burgess S, Small DS, Thompson SG. A review of instrumental variable estimators
  for Mendelian randomization. {\it Statistical Methods in Medical Research.}
  2017\string;26(5)\string:2333-2355.
\newblock \href {\doibase 10.1177/0962280215597579} {doi:
  10.1177/0962280215597579}

\bibitem{Didelez2007}
Didelez V, Sheehan N. Mendelian randomization as an instrumental variable
  approach to causal inference. {\it Statistical Methods in Medical Research.}
  2007\string;16(4)\string:309-330.
\newblock \href {\doibase 10.1177/09622802060Didel77743} {doi:
  10.1177/09622802060Didel77743}

\bibitem{Davies2015}
Davies NM, Hinke Kessler~Scholder vS, Farbmacher H, Burgess S, Windmeijer F,
  Smith GD. The many weak instruments problem and Mendelian randomization. {\it
  Statistics in Medicine.} 2015\string;34(3)\string:454-468.
\newblock \href {\doibase https://doi.org/10.1002/sim.6358} {doi:
  https://doi.org/10.1002/sim.6358}

\bibitem{Chao2005}
Chao JC, Swanson NR. Consistent Estimation with a Large Number of Weak
  Instruments. {\it Econometrica.} 2005\string;73(5)\string:1673-1692.
\newblock \href {\doibase https://doi.org/10.1111/j.1468-0262.2005.00632.x}
  {doi: https://doi.org/10.1111/j.1468-0262.2005.00632.x}

\bibitem{Bound1995}
John~Bound DAJ, Baker RM. Problems with Instrumental Variables Estimation when
  the Correlation between the Instruments and the Endogenous Explanatory
  Variable is Weak. {\it Journal of the American Statistical Association.}
  1995\string;90(430)\string:443-450.
\newblock \href {\doibase 10.1080/01621459.1995.10476536} {doi:
  10.1080/01621459.1995.10476536}

\bibitem{Hemani2018}
Hemani G, Bowden J, Davey~Smith G. {Evaluating the potential role of pleiotropy
  in Mendelian randomization studies}. {\it Human Molecular Genetics.}
  2018\string;27(R2)\string:R195-R208.
\newblock \href {\doibase 10.1093/hmg/ddy163} {doi: 10.1093/hmg/ddy163}

\bibitem{Verbanck2018}
Verbanck M, Chen CY, Neale B, Do R. Publisher Correction: Detection of
  widespread horizontal pleiotropy in causal relationships inferred from
  Mendelian randomization between complex traits and diseases. {\it Nature
  Genetics.} 2018\string;50(8)\string:1196.
\newblock \href {\doibase 10.1038/s41588-018-0164-2} {doi:
  10.1038/s41588-018-0164-2}

\bibitem{Tchetgen2021}
Tchetgen ET, Sun B, Walter S. {The GENIUS Approach to Robust Mendelian
  Randomization Inference}. {\it Statistical Science.}
  2021\string;36(3)\string:443 -- 464.
\newblock \href {\doibase 10.1214/20-STS802} {doi: 10.1214/20-STS802}

\bibitem{Pacini2016}
Pacini D, Windmeijer F. Robust inference for the Two-Sample 2SLS estimator.
  {\it Economics Letters.} 2016\string;146\string:50-54.
\newblock \href {\doibase https://doi.org/10.1016/j.econlet.2016.06.033} {doi:
  https://doi.org/10.1016/j.econlet.2016.06.033}

\bibitem{Guo2018}
Guo Z, Kang H, Tony~Cai T, Small DS. {Confidence Intervals for Causal Effects
  with Invalid Instruments by Using Two-Stage Hard Thresholding with Voting}.
  {\it Journal of the Royal Statistical Society Series B: Statistical
  Methodology.} 2018\string;80(4)\string:793-815.
\newblock \href {\doibase 10.1111/rssb.12275} {doi: 10.1111/rssb.12275}

\bibitem{Kang2016}
Hyunseung~Kang TTC, Small DS. Instrumental Variables Estimation With Some
  Invalid Instruments and its Application to Mendelian Randomization. {\it
  Journal of the American Statistical Association.}
  2016\string;111(513)\string:132-144.
\newblock \href {\doibase 10.1080/01621459.2014.994705} {doi:
  10.1080/01621459.2014.994705}

\bibitem{Sanderson2022}
Sanderson E, Glymour MM, Holmes MV, et al. Mendelian randomization. {\it Nature
  Reviews Methods Primers.} 2022\string;2(1)\string:6.
\newblock \href {\doibase 10.1038/s43586-021-00092-5} {doi:
  10.1038/s43586-021-00092-5}

\bibitem{Burgess2013}
Burgess S, Butterworth A, Thompson SG. Mendelian Randomization Analysis With
  Multiple Genetic Variants Using Summarized Data. {\it Genetic Epidemiology.}
  2013\string;37(7)\string:658-665.
\newblock \href {\doibase https://doi.org/10.1002/gepi.21758} {doi:
  https://doi.org/10.1002/gepi.21758}

\bibitem{Pierce2013}
Pierce BL, Burgess S. {Efficient Design for Mendelian Randomization Studies:
  Subsample and 2-Sample Instrumental Variable Estimators}. {\it American
  Journal of Epidemiology.} 2013\string;178(7)\string:1177-1184.
\newblock \href {\doibase 10.1093/aje/kwt084} {doi: 10.1093/aje/kwt084}

\bibitem{Bowden2017}
Bowden J, Del Greco M F, Minelli C, Davey~Smith G, Sheehan N, Thompson J. A
  framework for the investigation of pleiotropy in two-sample summary data
  Mendelian randomization. {\it Statistics in Medicine.}
  2017\string;36(11)\string:1783-1802.
\newblock \href {\doibase https://doi.org/10.1002/sim.7221} {doi:
  https://doi.org/10.1002/sim.7221}

\bibitem{Ye2021}
Ye T, Shao J, Kang H. {Debiased inverse-variance weighted estimator in
  two-sample summary-data Mendelian randomization}. {\it The Annals of
  Statistics.} 2021\string;49(4)\string:2079 -- 2100.
\newblock \href {\doibase 10.1214/20-AOS2027} {doi: 10.1214/20-AOS2027}

\bibitem{Xinwei2023}
Ma X, Wang J, Wu C. {Breaking the winner’s curse in Mendelian randomization:
  Rerandomized inverse variance weighted estimator}. {\it The Annals of
  Statistics.} 2023\string;51(1)\string:211 -- 232.
\newblock \href {\doibase 10.1214/22-AOS2247} {doi: 10.1214/22-AOS2247}

\bibitem{Xu2023}
Xu S, Wang P, Fung WK, Liu Z. A novel penalized inverse-variance weighted
  estimator for Mendelian randomization with applications to COVID-19 outcomes.
  {\it Biometrics.} 2023\string;79(3)\string:2184-2195.
\newblock \href {\doibase https://doi.org/10.1111/biom.13732} {doi:
  https://doi.org/10.1111/biom.13732}

\bibitem{Purcell2007}
Purcell S, Neale B, Todd-Brown K, et al. PLINK: a tool set for whole-genome
  association and population-based linkage analyses. {\it The American Journal
  of Human Genetics.} 2007\string;81(3)\string:559-75.
\newblock \href {\doibase 10.1086/519795} {doi: 10.1086/519795}

\bibitem{Bowden2016}
Bowden J, Davey~Smith G, Haycock PC, Burgess S. Consistent Estimation in
  Mendelian Randomization with Some Invalid Instruments Using a Weighted Median
  Estimator. {\it Genetic Epidemiology.} 2016\string;40(4)\string:304-314.
\newblock \href {\doibase https://doi.org/10.1002/gepi.21965} {doi:
  https://doi.org/10.1002/gepi.21965}

\bibitem{Bowden2015}
Bowden J, Davey~Smith G, Burgess S. {Mendelian randomization with invalid
  instruments: effect estimation and bias detection through Egger regression}.
  {\it International Journal of Epidemiology.}
  2015\string;44(2)\string:512-525.
\newblock \href {\doibase 10.1093/ije/dyv080} {doi: 10.1093/ije/dyv080}

\bibitem{Ziaeian2016}
Ziaeian B, Fonarow GC. Epidemiology and aetiology of heart failure. {\it Nature
  Reviews Cardiology.} 2016\string;13(6)\string:368-78.
\newblock \href {\doibase 10.1038/nrcardio.2016.25} {doi:
  10.1038/nrcardio.2016.25}

\bibitem{Paul2022}
Heidenreich PA, Bozkurt B, Aguilar D, et al. 2022 AHA/ACC/HFSA Guideline for
  the Management of Heart Failure. {\it Journal of the American College of
  Cardiology.} 2022\string;79(17)\string:e263-e421.
\newblock \href {\doibase 10.1016/j.jacc.2021.12.012} {doi:
  10.1016/j.jacc.2021.12.012}

\bibitem{Zhu2018}
Zhu Z, Zheng Z, Zhang F, et al. Causal associations between risk factors and
  common diseases inferred from GWAS summary data. {\it Nature Communications.}
  2018\string;9(1)\string:1--12.
\newblock \href {\doibase 10.1038/s41467-017-02317-2} {doi:
  10.1038/s41467-017-02317-2}

\bibitem{Shah2020}
Shah S, Henry A, Roselli C, et al. Genome-wide association and Mendelian
  randomisation analysis provide insights into the pathogenesis of heart
  failure. {\it Nature Communications.} 2020\string;11(1)\string:163.
\newblock \href {\doibase 10.1038/s41467-019-13690-5} {doi:
  10.1038/s41467-019-13690-5}

\end{thebibliography}

\bmsection*{Supporting information}

Web appendices, Tables, and Figures can be found online in the Supporting Information at the end of this article. The R code for the mdIVW method is publicly available at \href{https://github.com/YoupengSU/mdIVW}{https://github.com/YoupengSU/mdIVW}

\begin{table*}[ht] 
	\centering
	\caption{Comparison of the mdIVW estimator with other commonly used MR methods.
		The true causal effect $\beta_0=0.5$. 
		No horizontal pleiotropy exists ($\tau_0=0$) and no IV selection is conducted.
		The simulation is based on 20,000 repetitions.
		Bias (\%): bias divided by $\beta_0$; 
		SE: average of standard error; 
		MSE: mean squared error;
		CP: coverage probability of the 95\% confidence interval. \label{tab1}}%
	\begin{tabular*}{\textwidth}{@{\extracolsep\fill}lcrrrc@{\extracolsep\fill}}
		\toprule \textbf{$\hat{\psi}$} & \textbf{Method}  & \textbf{Bias (\%)}   & \textbf{SE} & \textbf{MSE} & \textbf{CP} \\
		\midrule
		8.13 & IVW	    &-79.47	&0.053	&0.161	 &0.000 \\
		& MR-Median		&-63.77	&0.090	&0.108	 &0.040 \\
		& MR-Egger	  	&-65.06 &0.082	&0.113	 &0.023 \\
		& MR-RAPS   	&18.76	&27.323	&393.416 &0.931 \\
		& dIVW	      	&4.90	&0.296	&0.090 	 &0.962 \\
		& mdIVW     	&-0.25	&0.273	&0.076	 &0.953 \\
		\midrule
		17.85& IVW      &-63.87  &0.048 &0.104	&0.000 \\
		& MR-Median		&-44.22	 &0.084	&0.054	&0.220 \\
		& MR-Egger	  	&-43.83	 &0.070	&0.053	&0.126 \\
		& MR-RAPS	  	&0.76	 &0.139 &0.020	&0.947 \\
		& dIVW     		&1.24	 &0.141	&0.020	&0.951 \\
		& mdIVW	    	&0.19	 &0.139 &0.020	&0.950 \\
		\midrule
		27.89& IVW      &-53.16  &0.044 &0.073	&0.000 \\
		& MR-Median     &-35.11  &0.074 &0.035	&0.326 \\
		& MR-Egger	 	&-32.11  &0.063	&0.030	&0.276 \\
		& MR-RAPS	  	&0.23	 &0.097	&0.009	&0.950 \\
		& dIVW	      	&0.41	 &0.098	&0.009	&0.952 \\
		& mdIVW	    	&-0.09	 &0.097	&0.009	&0.951 \\
		
		\bottomrule
	\end{tabular*}
\end{table*}

\begin{table*}[ht] 
	\centering
	\caption{Comparison of the mdIVW with other commonly used MR methods.
		The true causal effect $\beta_0=0.5$. 
		No horizontal pleiotropy exists ($\tau_0=0$). 
		The IV selection threshold $\delta=\sqrt{2{\rm log} p}$.
		The simulation is based $20,000$ repetitions.
		Bias (\%): bias divided by $\beta_0$; 
		SE: average of standard error; 
		MSE: mean squared error;
		CP: coverage probability of the 95\% confidence interval. \label{tab2}}%
	\begin{tabular*}{\textwidth}{@{\extracolsep\fill}lcrrrc@{\extracolsep\fill}}
		\toprule \textbf{$\hat{\psi}_\delta$} & \textbf{Method}  & \textbf{Bias (\%)}   & \textbf{SE} & \textbf{MSE} & \textbf{CP} \\
		\midrule
		7.22 & IVW	    &-2.82	&0.108	&0.011	 &0.954 \\
		& MR-Median		&-4.44  &0.138 	&0.016	 &0.971 \\
		& MR-Egger	  	&-22.29  &0.340	&0.122	 &0.940 \\
		& MR-RAPS   	&0.21	&0.109	&0.012   &0.956 \\
		& dIVW	      	&0.50	&0.106	&0.011 	 &0.952 \\
		& mdIVW     	&-0.12	&0.105	&0.011	 &0.951 \\
		\midrule
		10.21& IVW      &-3.30   &0.094 &0.008	&0.952 \\
		& MR-Median		&-5.14	 &0.125	&0.013	&0.969 \\
		& MR-Egger	  	&-24.72	 &0.278	&0.087	&0.930 \\
		& MR-RAPS	  	&0.18	 &0.095 &0.009	&0.951 \\
		& dIVW     		&0.42	 &0.093	&0.009	&0.951 \\
		& mdIVW	    	&-0.05	 &0.093 &0.009	&0.951 \\
		\midrule
		16.03& IVW      &-4.02  &0.080  &0.006	&0.947 \\
		& MR-Median     & -6.33 &0.111  &0.011	&0.965 \\
		& MR-Egger	 	&-27.96 &0.220	&0.066	&0.902 \\
		& MR-RAPS	  	&0.21	&0.083	&0.007	&0.952 \\
		& dIVW	      	&0.36	&0.081	&0.007	&0.952 \\
		& mdIVW	    	&0.01	&0.081	&0.007	&0.952 \\
		
		\bottomrule
	\end{tabular*}
\end{table*}
\clearpage

\begin{table*}[ht] 
	\centering
	\caption{Estimated causal effects ($\hat{\beta}$) and estimated standard errors (SEs) of coronary artery disease (CAD) and dyslipidemia on the risk of heart failure. \label{tab3}}
	\begin{tabular*}{500pt}{@{\extracolsep\fill}llcccc@{\extracolsep\fill}}
		\toprule
		\multicolumn{2}{c}{}&\multicolumn{2}{@{}c@{}}{\textbf{No IV selection}} & \multicolumn{2}{@{}c@{}} {\textbf{IV selection with} $\boldsymbol{\delta=\sqrt{2{\rm log} p}}$} \\ \cmidrule{3-4}\cmidrule{5-6}
		\textbf{Exposure} & \textbf{Method}  & ${\hat{\psi}}$ & $\hat{\beta}$ (SE) & ${\hat{\psi}}_\delta$  & $\hat{\beta}$ (SE)   \\
		\midrule
		CAD	 & IVW	    & 5.05	& 0.098 (0.013)	& 4.59	& 0.336 (0.061) \\
		& MR-Median	    &       & 0.033 (0.020)	&		& 0.054 (0.088) \\
		& MR-Egger		&       & 0.090 (0.019)	&		& 0.439 (0.084) \\
		& MR-RAPS		&       & 0.890 (0.147)	&	    & 0.903 (0.161) \\
		& dIVW		    &       & 1.086 (0.359)	&	    & 0.997 (0.316) \\
		& mdIVW		    &       & 0.982 (0.243)	&		& 0.909 (0.248) \\
		\midrule
		Dyslipidemia &IVW & 30.90  & 0.077 (0.011)		& 30.36	  & 0.179 (0.023) \\
		&MR-Median	      &	       & 0.104 (0.022)		&         &	0.201 (0.031) \\
		&MR-Egger	 	  &        & 0.088 (0.016)		&	      & 0.187 (0.029) \\
		&MR-RAPS		  &        & 0.236 (0.025)		&	      & 0.217 (0.021) \\
		&dIVW		      &        & 0.195 (0.029)		&	      & 0.203 (0.021) \\
		&mdIVW		      &        & 0.194 (0.029)		&	      & 0.203 (0.021) \\
		\bottomrule
	\end{tabular*}
\end{table*}

\clearpage

\begin{figure}[htb]
	\centering
	\includegraphics[width=1\textwidth]{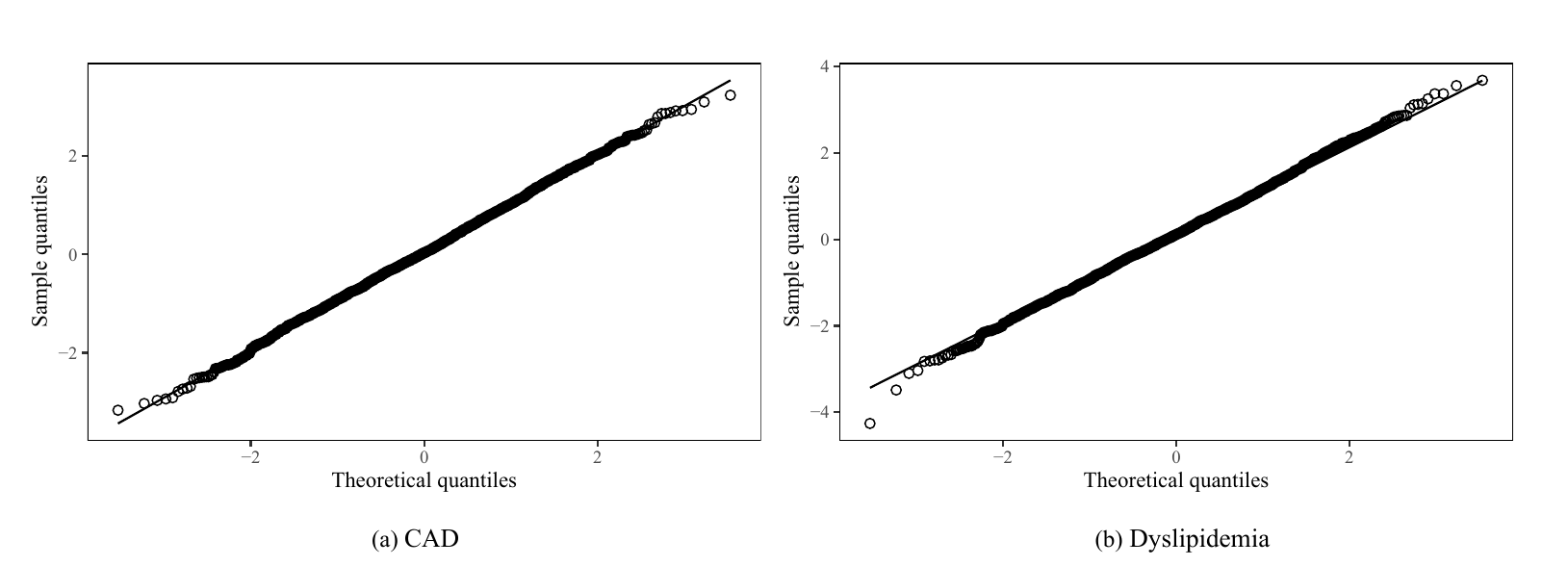}
	\caption{
		Quantile-Quantile plots of the standardized residuals against a standard normal distribution.}
	\label{fig4}
\end{figure}
\clearpage



\section*{Supplementary material}

# Contents

# Web Appendix A: Proof of Theorem 1

In the following proofs, for two sequences of real numbers $a_n$ and $b_n$ indexed by $n$, we write $a_n = O(b_n)$ if there exists a constant $c$ such that $|a_n| \leq cb_n$ for all $n$, $a_n = o(b_n)$ if $a_n/b_n \to 0$ as $n \to \infty$, and $a_n = \Theta(b_n)$ if there exists a constant $c$ such that $c^{-1}b_n \leq |a_n| \leq cb_n$. We use $\xrightarrow{P}$ and $\xrightarrow{D}$ to denote convergence in probability and distribution separately. For random variables $X$ and $Y$, we denote $X = o_p(Y)$ if $X/Y \xrightarrow{P} 0$, and $X = O_p(Y)$ if $X/Y$ is bounded in probability.

The following proofs hold for both cases with ($\tau_0 \neq 0$) and without ($\tau_0 = 0$) balanced horizontal pleiotropy. In fact, no horizontal pleiotropy is a special case of balanced horizontal pleiotropy with $\tau_0 = 0$.

## A.1 Bias of the dIVW Estimator

The dIVW estimator is written as

$$\hat{\beta}_{\text{dIVW}} = \frac{\hat{\theta}_1}{\hat{\theta}_2} = \frac{\sum_{j=1}^p \hat{\sigma}_{\hat{\Gamma}_j}^{-2} \hat{\gamma}_j \hat{\Gamma}_j}{\sum_{j=1}^p \hat{\sigma}_{\hat{\Gamma}_j}^{-2} \left(\hat{\gamma}_j^2 - \hat{\sigma}_{\hat{\gamma}_j}^{-2}\right)}.$$

Let $\hat{x}_1 = \hat{\theta}_1/\theta_1 - 1$ and $\hat{x}_2 = \hat{\theta}_2/\theta_2 - 1$, where $\theta_1 = E(\hat{\theta}_1) = \beta_0 \sum_{j=1}^p \sigma_{\hat{\Gamma}_j}^{-2} \gamma_j^2 = \Theta(\kappa p)$ and $\theta_2 = E(\hat{\theta}_2) = \sum_{j=1}^p \sigma_{\hat{\Gamma}_j}^{-2} \gamma_j^2 = \Theta(\kappa p)$ under Assumption 2. Through Taylor series expansion, we have

$$\hat{\beta}_{\text{dIVW}} = \frac{\theta_1(\hat{x}_1 + 1)}{\theta_2(\hat{x}_2 + 1)} = \beta_0\left(1 + \hat{x}_1 - \hat{x}_2 - \hat{x}_1\hat{x}_2 + \hat{x}_2^2\right) + o_p(\varphi),$$

where $\varphi = 1/\kappa p + 1/\kappa^2 p \to 0$ as the effective sample size $\psi = \kappa\sqrt{p} \to \infty$. Then the bias of $\hat{\beta}_{\text{dIVW}}$ is

$$E\left(\hat{\beta}_{\text{dIVW}} - \beta_0\right) = \beta_0\left(\frac{v_2}{\theta_2^2} - \frac{v_{12}}{\theta_1\theta_2}\right) + o(\varphi) = O(\varphi),$$

where $v_2 = \text{Var}(\hat{\theta}_2) = 2\sum_{j=1}^p \sigma_{\hat{\Gamma}_j}^{-4}\left(2\sigma_{\hat{\gamma}_j}^2 \gamma_j^2 + \sigma_{\hat{\gamma}_j}^4\right) = \Theta(\kappa p + p)$, and $v_{12} = \text{Cov}(\hat{\theta}_1, \hat{\theta}_2) = 2\beta_0 \sum_{j=1}^p \sigma_{\hat{\Gamma}_j}^{-4} \sigma_{\hat{\gamma}_j}^2 \gamma_j^2 = \Theta(\kappa p)$ by Assumption 2.

3

## A.2 Bias of the mdIVW Estimator

The mdIVW estimator is written as

$$\widehat{\beta}_{\text{mdIVW}} = \frac{\widehat{\theta}_1}{\widehat{\theta}_2}\left(1 - \frac{\widehat{v}_2}{\widehat{\theta}_2^2}\right) + \frac{\widehat{v}_{12}}{\widehat{\theta}_2^2}.$$

Through Taylor series expansion, we have

$$\begin{aligned}\widehat{\beta}_{\text{mdIVW}} &= \frac{\theta_1(\widehat{x}_1 + 1)}{\theta_2(\widehat{x}_2 + 1)}\left(1 - \frac{\widehat{v}_2}{\widehat{\theta}_2^2}\right) + \frac{\widehat{v}_{12}}{\widehat{\theta}_2^2} \\ &= \beta_0\left(1 + \widehat{x}_1 - \widehat{x}_2 - \widehat{x}_1\widehat{x}_2 + \widehat{x}_2^2 + \widehat{x}_1\widehat{x}_2^2 - \widehat{x}_2^3 - \widehat{x}_1\widehat{x}_2^3 - \frac{\widehat{v}_2}{\widehat{\theta}_2^2} - \widehat{x}_1\frac{\widehat{v}_2}{\widehat{\theta}_2^2} + \widehat{x}_2\frac{\widehat{v}_2}{\widehat{\theta}_2^2}\right) + \frac{\widehat{v}_{12}}{\widehat{\theta}_2^2} + O_p(\varphi^2)\end{aligned}$$

After some algebra, the expectations of $\widehat{x}_1\widehat{x}_2^2$, $\widehat{x}_2^3$ and $\widehat{x}_1\widehat{x}_2^3$ are of order $O(\varphi^2)$. Then the bias of $\widehat{\beta}_{\text{mdIVW}}$ is

$$E\left(\widehat{\beta}_{\text{mdIVW}} - \beta_0\right) = \beta_0 \underbrace{E\left(\frac{v_2}{\theta_2^2} - \frac{\widehat{v}_2}{\widehat{\theta}_2^2}\right)}_{\mathbf{A}} + \underbrace{E\left(\frac{\widehat{v}_{12}}{\widehat{\theta}_2^2} - \frac{v_{12}}{\theta_2^2}\right)}_{\mathbf{B}} + \underbrace{E\left(-\widehat{x}_1\frac{\widehat{v}_2}{\widehat{\theta}_2^2} + \widehat{x}_2\frac{\widehat{v}_2}{\widehat{\theta}_2^2}\right)}_{\mathbf{C}} + O\left(\varphi^2\right).$$

For **A**, let $\widehat{y}_2 = \widehat{v}_2/v_2 - 1$, Through Taylor series expansion we have

$$\frac{\widehat{v}_2}{\widehat{\theta}_2^2} = \frac{v_2(\widehat{y}_2 + 1)}{\theta_2^2(\widehat{x}_2 + 1)^2} = \frac{v_2}{\theta_2^2}\left\{1 - 2\widehat{x}_2 + 3\widehat{x}_2^2 + \widehat{y}_2 - 2\widehat{x}_2\widehat{y}_2 + o_p(\varphi)\right\}.$$

Then we have

$$E\left(\frac{v_2}{\theta_2^2} - \frac{\widehat{v}_2}{\widehat{\theta}_2^2}\right) = -\frac{3v_2^2}{\theta_2^4} + \frac{2\text{Cov}(\widehat{\theta}_2, \widehat{v}_2)}{\theta_2^3} + o\left(\varphi^2\right) = O\left(\varphi^2\right), \tag{1}$$

where $\text{Cov}(\widehat{\theta}_2, \widehat{v}_2) = 8\sum_{j=1}^{p}\sigma_{\widehat{\Gamma}_j}^{-6}\sigma_{\widehat{\gamma}_j}^4\left(\sigma_{\widehat{\gamma}_j}^2 + 2\gamma_j^2\right) = \Theta(kp + p)$.

For **B**, let $\widehat{y}_{12} = \widehat{v}_{12}/v_{12} - 1$. Through Taylor series expansion we have

$$\frac{\widehat{v}_{12}}{\widehat{\theta}_2^2} = \frac{v_{12}(\widehat{y}_{12} + 1)}{\theta_2^2(\widehat{x}_2 + 1)^2} = \frac{v_{12}}{\theta_2^2}\left\{1 - 2\widehat{x}_2 + 3\widehat{x}_2^2 + \widehat{y}_{12} - 2\widehat{x}_2\widehat{y}_{12} + o_p(\varphi)\right\}.$$

Then we have

$$E\left(\frac{\widehat{v}_{12}}{\widehat{\theta}_2^2} - \frac{v_{12}}{\theta_2^2}\right) = \frac{3v_2v_{12}}{\theta_2^4} - \frac{2\text{Cov}(\widehat{\theta}_2, \widehat{v}_{12})}{\theta_2^3} + o\left(\varphi^2\right) = O\left(\varphi^2\right), \tag{2}$$

where $\text{Cov}\left(\widehat{\theta}_2, \widehat{v}_{12}\right) = 4\beta_0 \sum_{j=1}^{p} \sigma_{\widehat{\Gamma}_j}^{-6} \sigma_{\widehat{\gamma}_j}^{4} \gamma_j^2 = \Theta\left(kp\right)$.

For **C**, through Taylor series expansion we have

$$\widehat{x}_1 \frac{\widehat{v}_2}{\widehat{\theta}_2^2} = \widehat{x}_1 \frac{v_2}{\theta_2^2} \left\{1 - 2\widehat{x}_2 + 3\widehat{x}_2^2 + \widehat{y}_2 - 2\widehat{x}_2\widehat{y}_2 + o_p(\varphi)\right\} = \frac{v_2}{\theta_2^2} \left(\widehat{x}_1 - 2\widehat{x}_1\widehat{x}_2 + \widehat{x}_1\widehat{y}_2 + o_p(\varphi)\right),$$

$$\widehat{x}_2 \frac{\widehat{v}_2}{\widehat{\theta}_2^2} = \widehat{x}_2 \frac{v_2}{\theta_2^2} \left\{1 - 2\widehat{x}_2 + 3\widehat{x}_2^2 + \widehat{y}_2 - 2\widehat{x}_2\widehat{y}_2 + o_p(\varphi)\right\} = \frac{v_2}{\theta_2^2} \left(\widehat{x}_2 - 2\widehat{x}_2^2 + \widehat{x}_2\widehat{y}_2 + o_p(\varphi)\right).$$

And their expectations are

$$E\left(\widehat{x}_1 \frac{\widehat{v}_2}{\widehat{\theta}_2^2}\right) = \frac{-2v_{12}v_2}{\theta_1 \theta_2^3} + \frac{\text{Cov}(\widehat{\theta}_1, \widehat{v}_2)}{\theta_1 \theta_2^2} + o\left(\varphi^2\right) = O\left(\varphi^2\right), \tag{3}$$

$$E\left(\widehat{x}_2 \frac{\widehat{v}_2}{\widehat{\theta}_2^2}\right) = \frac{-2v_2^2}{\theta_2^4} + \frac{\text{Cov}(\widehat{\theta}_2, \widehat{v}_2)}{\theta_2^3} + o\left(\varphi^2\right) = O\left(\varphi^2\right), \tag{4}$$

where $\text{Cov}\left(\widehat{\theta}_1, \widehat{v}_2\right) = 8\beta_0 \sum_{j=1}^{p} \sigma_{\widehat{\Gamma}_j}^{-6} \sigma_{\widehat{\gamma}_j}^{4} \gamma_j^2 = \Theta\left(kp\right)$. Combining Equations (1-4), we have

$$E\left(\widehat{\beta}_{\text{mdIVW}} - \beta_0\right) = O\left(\varphi^2\right).$$

## A.3 the Variance Difference between $\widehat{\beta}_{\text{dIVW}}$ and $\widehat{\beta}_{\text{mdIVW}}$

Since $\widehat{\beta}_{\text{mdIVW}} = \widehat{\beta}_{\text{dIVW}} + \frac{\widehat{v}_{12}}{\widehat{\theta}_2^2} - \frac{\widehat{\theta}_1 \widehat{v}_2}{\widehat{\theta}_2^3}$, we have

$$\begin{aligned}
\text{Var}\left(\widehat{\beta}_{\text{mdIVW}}\right) &= \text{Var}\left(\widehat{\beta}_{\text{dIVW}}\right) + \text{Var}\left(\frac{\widehat{v}_{12}}{\widehat{\theta}_2^2} - \frac{\widehat{\theta}_1 \widehat{v}_2}{\widehat{\theta}_2^3}\right) + 2\text{Cov}\left(\widehat{\beta}_{\text{dIVW}} - \beta_0, \frac{\widehat{v}_{12}}{\widehat{\theta}_2^2} - \frac{\widehat{\theta}_1 \widehat{v}_2}{\widehat{\theta}_2^3}\right) \\
&= \text{Var}\left(\widehat{\beta}_{\text{dIVW}}\right) + 2\text{Cov}\left(\widehat{\beta}_{\text{dIVW}} - \beta_0, \frac{\widehat{v}_{12}}{\widehat{\theta}_2^2} - \frac{\widehat{\theta}_1 \widehat{v}_2}{\widehat{\theta}_2^3}\right) + o\left(\varphi^2\right).
\end{aligned}$$

By Taylor series expansion, we have

$$\widehat{\beta}_{\text{dIVW}} - \beta_0 = \beta_0 \left(\widehat{x}_1 - \widehat{x}_2\right) + O_p(\varphi),$$

$$\frac{\widehat{v}_{12}}{\widehat{\theta}_2^2} - \frac{\widehat{\theta}_1 \widehat{v}_2}{\widehat{\theta}_2^3} = \left(\frac{v_{12}}{\theta_2^2} - \frac{\theta_1 v_2}{\theta_2^3}\right) - \frac{\theta_1 v_2}{\theta_2^3} \widehat{x}_1 + \left(\frac{3\theta_1 v_2}{\theta_2^3} - \frac{2v_{12}}{\theta_2^2}\right)\widehat{x}_2 + \frac{v_{12}}{\theta_2^2} \widehat{y}_{12} - \frac{\theta_1 v_2}{\theta_2^3} \widehat{y}_2 + O_p\left(\varphi^2\right).$$

5

Therefore, we have

$$
\begin{aligned}
&\mathrm{Cov}\left(\widehat{\beta}_{\mathrm{dIVW}} - \beta_0, \frac{\widehat{v}_{12}}{\widehat{\theta}_2^2} - \frac{\widehat{\theta}_1 \widehat{v}_2}{\widehat{\theta}_2^3}\right) \\
&= \mathrm{Cov}\left(\beta_0\left(\widehat{x}_1 - \widehat{x}_2\right), -\frac{\theta_1 v_2}{\theta_2^3}\widehat{x}_1 + \left(\frac{3\theta_1 v_2}{\theta_2^3} - \frac{2v_{12}}{\theta_2^2}\right)\widehat{x}_2 + \frac{v_{12}}{\theta_2^2}\widehat{y}_{12} - \frac{\theta_1 v_2}{\theta_2^3}\widehat{y}_2\right) + o\left(\varphi^2\right) \\
&= \beta_0 \left\{ -\frac{\theta_1 v_2}{\theta_2^3}\mathrm{Var}\left(\widehat{x}_1\right) + \left(\frac{4\theta_1 v_2}{\theta_2^3} - \frac{2v_{12}}{\theta_2^2}\right)\mathrm{Cov}\left(\widehat{x}_1, \widehat{x}_2\right) - \left(\frac{3\theta_1 v_2}{\theta_2^3} - \frac{2v_{12}}{\theta_2^3}\right)\mathrm{Var}\left(\widehat{x}_2\right) + \frac{v_{12}}{\theta_2^2}\mathrm{Cov}\left(\widehat{x}_1, \widehat{y}_{12}\right) \right. \\
&\qquad \left. -\frac{\theta_1 v_2}{\theta_2^3}\mathrm{Cov}\left(\widehat{x}_1, \widehat{y}_2\right) - \frac{v_{12}}{\theta_2^2}\mathrm{Cov}\left(\widehat{x}_2, \widehat{y}_{12}\right) + \frac{\theta_1 v_2}{\theta_2^3}\mathrm{Cov}\left(\widehat{x}_2, \widehat{y}_2\right) \right\} + o\left(\varphi^2\right) \\
&= -\frac{\beta_0^2}{\theta_2^4}\left\{\frac{v_1 v_2}{\beta_0^2} - \frac{6 v_{12} v_2}{\beta_0} + \frac{2 v_{12}^2}{\beta_0^2} + 3 v_2^2 - \frac{\theta_2 \mathrm{Cov}(\widehat{\theta}_1, \widehat{v}_{12})}{\beta_0^2} + \frac{\theta_2 \mathrm{Cov}(\widehat{\theta}_1, \widehat{v}_2)}{\beta_0} + \frac{\theta_2 \mathrm{Cov}(\widehat{\theta}_2, \widehat{v}_{12})}{\beta_0} \right. \\
&\qquad \left. - \theta_2 \mathrm{Cov}\left(\widehat{\theta}_2, \widehat{v}_2\right) \right\} + o\left(\varphi^2\right) \\
&= -\frac{\beta_0^2}{\theta_2^4}\Delta + o\left(\varphi^2\right),
\end{aligned}
$$

where $\mathrm{Cov}\left(\widehat{\theta}_1, \widehat{v}_{12}\right) = 2\sum_{j=1}^{p}\sigma_{\widehat{\Gamma}_j}^{-6}\sigma_{\widehat{\gamma}_j}^{2}\left\{\beta_0^2 \sigma_{\widehat{\gamma}_j}^2 \gamma_j^2 + \left(\sigma_{\widehat{\Gamma}_j}^2 + \tau_0^2\right)\left(\gamma_j^2 + \sigma_{\widehat{\gamma}_j}^2\right)\right\} = \Theta(\kappa p + p)$,

$$
v_1 = \mathrm{Var}\left(\widehat{\theta}_1\right) = \sum_{j=1}^{p}\sigma_{\widehat{\Gamma}_j}^{-4}\left\{\beta_0^2 \sigma_{\widehat{\gamma}_j}^2 \gamma_j^2 + \left(\sigma_{\widehat{\Gamma}_j}^2 + \tau_0^2\right)\left(\gamma_j^2 + \sigma_{\widehat{\gamma}_j}^2\right)\right\} = \Theta(\kappa p + p),
$$

and

$$
\Delta = \frac{v_1 v_2}{\beta_0^2} - \frac{6 v_{12} v_2}{\beta_0} + \frac{2 v_{12}^2}{\beta_0^2} + 3v_2^2 - \theta_2 \sum_{j=1}^{p}\sigma_{\widehat{\Gamma}_j}^{-6}\sigma_{\widehat{\gamma}_j}^{6}\left(6\sigma_{\widehat{\gamma}_j}^{-2}\gamma_j^2 + 8\right) - 2\theta_2 \beta_0^{-2}\sum_{j=1}^{p}\sigma_{\widehat{\Gamma}_j}^{-4}\sigma_{\widehat{\gamma}_j}^{4}(\sigma_{\widehat{\gamma}_j}^{-2}\gamma_j^2 + 1)(\sigma_{\widehat{\Gamma}_j}^{-2}\tau_0^2 + 1)
$$

So, we have

$$
\mathrm{Var}\left(\widehat{\beta}_{\mathrm{dIVW}}\right) - \mathrm{Var}\left(\widehat{\beta}_{\mathrm{mdIVW}}\right) = \frac{2\beta_0^2}{\theta_2^4}\Delta + o\left(\varphi^2\right).
$$

# Web Appendix B: Proof of $\Delta > 0$

The following proofs hold for both cases with ($\tau_0 \neq 0$) and without ($\tau_0 = 0$) balanced horizontal pleiotropy. No horizontal pleiotropy is a special case of balanced horizontal pleiotropy with $\tau_0 = 0$.

Let $k_j = \gamma_j^2/\sigma_{\hat{\gamma}_j}^2$. After some algebra we have

$$\Delta = \frac{v_1 v_2}{\beta_0^2} - \frac{6 v_{12} v_2}{\beta_0} + \frac{2 v_{12}^2}{\beta_0^2} + 3v_2^2 - \theta_2 \sum_{j=1}^{p} \sigma_{\hat{\Gamma}_j}^{-6} \sigma_{\hat{\gamma}_j}^{6} (8 + 6k_j) - 2\beta_0^{-2}\theta_2 \sum_{j=1}^{p} \left( \sigma_{\hat{\Gamma}_j}^{-4} \sigma_{\hat{\gamma}_j}^{4} + \sigma_{\hat{\Gamma}_j}^{-6} \sigma_{\hat{\gamma}_j}^{4} \tau_0^2 \right)(1+k_j)$$

$$= \left\{ v_2 \sum_{j=1}^{p} \left\{ \sigma_{\hat{\Gamma}_j}^{-4} \sigma_{\hat{\gamma}_j}^{4} k_j + \beta_0^{-2} \sigma_{\hat{\Gamma}_j}^{-2} \sigma_{\hat{\gamma}_j}^{2} (1+k_j) \right\} + 6v_2 \sum_{j=1}^{p} \sigma_{\hat{\Gamma}_j}^{-4} \sigma_{\hat{\gamma}_j}^{4} + 8 \left( \sum_{j=1}^{p} \sigma_{\hat{\Gamma}_j}^{-4} \sigma_{\hat{\gamma}_j}^{4} k_j \right)^2 - \theta_2 \sum_{j=1}^{p} \sigma_{\hat{\Gamma}_j}^{-6} \sigma_{\hat{\gamma}_j}^{6} (8+6k_j) \right.$$

$$\left. - 2\beta_0^{-2}\theta_2 \sum_{j=1}^{p} \sigma_{\hat{\Gamma}_j}^{-4} \sigma_{\hat{\gamma}_j}^{4} (1+k_j) \right\} + \left( \beta_0^{-2} \tau_0^2 \right) \left\{ v_2 \sum_{j=1}^{p} \sigma_{\hat{\Gamma}_j}^{-4} \sigma_{\hat{\gamma}_j}^{2} (1+k_j) - 2\theta_2 \sum_{j=1}^{p} \sigma_{\hat{\Gamma}_j}^{-6} \sigma_{\hat{\gamma}_j}^{4} (1+k_j) \right\}$$

$$= \mathbf{D} + \mathbf{E},$$

where $\mathbf{D}$ represents the part of the curly brackets before the plus sign, and $\mathbf{E}$ represents the part after the plus sign. A sufficient condition for $\Delta > 0$ is $\mathbf{D} > 0$ and $\mathbf{E} \geqslant 0$. For $\mathbf{D}$, we have

$$\mathbf{D} > v_2 \sum_{j=1}^{p} \sigma_{\hat{\Gamma}_j}^{-4} \sigma_{\hat{\gamma}_j}^{4} k_j + \beta_0^{-2} \theta_2 v_2 + 6v_2 \sum_{j=1}^{p} \sigma_{\hat{\Gamma}_j}^{-4} \sigma_{\hat{\gamma}_j}^{4} + 8 \left( \sum_{j=1}^{p} \sigma_{\hat{\Gamma}_j}^{-4} \sigma_{\hat{\gamma}_j}^{4} k_j \right)^2 - \theta_2 \sum_{j=1}^{p} \sigma_{\hat{\Gamma}_j}^{-6} \sigma_{\hat{\gamma}_j}^{6} (8+6k_j)$$

$$- 2\beta_0^{-2}\theta_2 \sum_{j=1}^{p} \sigma_{\hat{\Gamma}_j}^{-4} \sigma_{\hat{\gamma}_j}^{4} (1+k_j)$$

$$> v_2 \sum_{j=1}^{p} \sigma_{\hat{\Gamma}_j}^{-4} \sigma_{\hat{\gamma}_j}^{4} k_j + 6v_2 \sum_{j=1}^{p} \sigma_{\hat{\Gamma}_j}^{-4} \sigma_{\hat{\gamma}_j}^{4} + 8 \left( \sum_{j=1}^{p} \sigma_{\hat{\Gamma}_j}^{-4} \sigma_{\hat{\gamma}_j}^{4} k_j \right)^2 - \theta_2 \sum_{j=1}^{p} \sigma_{\hat{\Gamma}_j}^{-6} \sigma_{\hat{\gamma}_j}^{6} (8+6k_j)$$

$$= 2 \sum_{j=1}^{p} \sigma_{\hat{\Gamma}_j}^{-4} \sigma_{\hat{\gamma}_j}^{4} (2k_j+3) \sum_{j=1}^{p} \sigma_{\hat{\Gamma}_j}^{-4} \sigma_{\hat{\gamma}_j}^{4} (3k_j+2) - 2 \sum_{j=1}^{p} \sigma_{\hat{\Gamma}_j}^{-2} \sigma_{\hat{\gamma}_j}^{2} k_j \sum_{j=1}^{p} \sigma_{\hat{\Gamma}_j}^{-6} \sigma_{\hat{\gamma}_j}^{6} (4+3k_j),$$

where we obtain the first inequality by omitting the term $\beta_0^{-2} v_2 \sum_{j=1}^{p} \sigma_{\hat{\Gamma}_j}^{-2} \sigma_{\hat{\gamma}_j}^{2}$ and the second inequality by using $v_2 > 2 \sum_{j=1}^{p} \sigma_{\hat{\Gamma}_j}^{-4} \sigma_{\hat{\gamma}_j}^{4} (1+k_j)$. Therefore, a sufficient condition of $\mathbf{D} > 0$ is

$$\sum_{j=1}^{p} \sigma_{\hat{\Gamma}_j}^{-4} \sigma_{\hat{\gamma}_j}^{4} (2k_j+3) \sum_{j=1}^{p} \sigma_{\hat{\Gamma}_j}^{-4} \sigma_{\hat{\gamma}_j}^{4} (3k_j+2) > \sum_{j=1}^{p} \sigma_{\hat{\Gamma}_j}^{-2} \sigma_{\hat{\gamma}_j}^{2} k_j \sum_{j=1}^{p} \sigma_{\hat{\Gamma}_j}^{-6} \sigma_{\hat{\gamma}_j}^{6} (4+3k_j) \qquad (5)$$

7

Since, $\hat{\gamma}_j$ and $\hat{\Gamma}_j$ are estimated form GWAS marginal linear models, their variances can be expressed as $\sigma^2_{\hat{\gamma}_j} = \text{Var}(X)(1-h_{X,j})/\{n_X\text{Var}(Z_j)\}$ and $\sigma^2_{\hat{\Gamma}_j} = \text{Var}(Y)(1-h_{Y,j})/\{n_Y Var(Z_j)\}$, where $h_{X,j}$ and $h_{Y,j}$ denote the variances of $X$ and $Y$ explained by $Z_j$. Let $c_j^2 = (1-h_{X,j})/(1-h_{Y,j})$, then the Inequality (5) can be simplified as

$$\sum_{j=1}^p c_j^4 (3+2k_j) \sum_{j=1}^p c_j^4 (2+3k_j) > \sum_{j=1}^p c_j^2 k_j \sum_{j=1}^p \{c_j^6 (4+3k_j)\}. \qquad (6)$$

For the left-hand side of Inequality (6), we have

$$\sum_{j=1}^p c_j^4 (2k_j+3) \sum_{j=1}^p c_j^4 (3k_j+2) > 6\sum_{j=1}^p c_j^4 k_j \sum_{j=1}^p c_j^4 k_j + 13\sum_{j=1}^p c_j^4 k_j \sum_{j=1}^p c_j^4. \qquad (7)$$

Let

$$\Delta_{1j} = c_j^2 - c_j^4 = \frac{(1-h_{X,j})(h_{X,j}-h_{Y,j})}{(1-h_{Y,j})^2},$$

$$\Delta_{2j} = c_j^4 - c_j^6 = \frac{(1-h_{X,j})^2(h_{Y,j}-h_{X,j})}{(1-h_{Y,j})^3}.$$

Then, for the right-hand side of Inequality (6), we have

$$\sum_{j=1}^p c_j^2 k_j \sum_{j=1}^p \{c_j^6 (4+3k_j)\} = 3\sum_{j=1}^p (c_j^4+\Delta_{1j})k_j \sum_{j=1}^p (c_j^4+\Delta_{2j})k_j + 4\sum_{j=1}^p (c_j^4+\Delta_{1j})k_j \sum_{j=1}^p (c_j^4+\Delta_{2j}). \qquad (8)$$

Using Inequality (7) and Equality (8), the inequality (5) holds when

$$6\sum_{j=1}^p c_j^4 k_j \sum_{j=1}^p c_j^4 k_j + 13\sum_{j=1}^p c_j^4 k_j \sum_{j=1}^p c_j^4 > 3\sum_{j=1}^p (c_j^4+\Delta_{1j})k_j \sum_{j=1}^p (c_j^4+\Delta_{2j})k_j + 4\sum_{j=1}^p (c_j^4+\Delta_{1j})k_j \sum_{j=1}^p (c_j^4+\Delta_{2j}). \qquad (9)$$

Notice that, since $\Delta_{1j}\Delta_{2j} < 0$, we have either $\Delta_{1j} < 0$ or $\Delta_{2j} < 0$. Therefore, the Inequality (9) holds when $6c_j^4 > 3(c_j^4+\Delta_{1j})$ and $6c_j^4 > 3(c_j^4+\Delta_{2j})$, that is

$$c_j^4 > \max(\Delta_{1j}, \Delta_{2j}),$$

which holds as long as $\max(h_{X,j}, h_{Y,j}) < 1/2$.

For **E**, if there is no balanced horizontal pleiotropy, then $\mathbf{E} = 0$. For $\mathbf{E} \geqslant 0$, it is sufficient to prove

$$v_2 \sum_{j=1}^{p} \sigma_{\widehat{\Gamma}_j}^{-4} \sigma_{\widehat{\gamma}_j}^{2} (1 + k_j) - 2\theta_2 \sum_{j=1}^{p} \sigma_{\widehat{\Gamma}_j}^{-6} \sigma_{\widehat{\gamma}_j}^{4} (1 + k_j) > 0.$$

Similar to the proof of $\mathbf{D} > 0$, a sufficient condition of $\mathbf{E} \geqslant 0$ is

$$\sum_{j=1}^{p} \sigma_{\widehat{\Gamma}_j}^{-4} \sigma_{\widehat{\gamma}_j}^{4} (2k_j + 1) \sum_{j=1}^{p} \sigma_{\widehat{\Gamma}_j}^{-4} \sigma_{\widehat{\gamma}_j}^{2} (1 + k_j) > \sum_{j=1}^{p} \sigma_{\widehat{\Gamma}_j}^{-2} \sigma_{\widehat{\gamma}_j}^{2} k_j \sum_{j=1}^{p} \sigma_{\widehat{\Gamma}_j}^{-6} \sigma_{\widehat{\gamma}_j}^{4} (1 + k_j),$$

which holds when

$$2\sum_{j=1}^{p} c_j^4 k_j \sum_{j=1}^{p} \sigma_{\widehat{\Gamma}_j}^{-2} c_j^2 (1 + k_j) > \sum_{j=1}^{p} c_j^2 k_j \sum_{j=1}^{p} \sigma_{\widehat{\Gamma}_j}^{-2} c_j^4 (1 + k_j). \tag{10}$$

Inequality (10) holds when $2c_j^4 > c_j^2$ and $c_j^2 > c_j^4$ or $c_j^4 > c_j^2$ and $2c_j^2 > c_j^4$, that is,

$$0.5 < c_j^2 < 2,$$

which holds as long as $\max(h_{X,j}, h_{Y,j}) < 1/2$.

Thus, $\text{Var}(\widehat{\beta}_{\text{mdIVW}})$ is asymptotically smaller than $\text{Var}(\widehat{\beta}_{\text{dIVW}})$ as long as the proportion of variation in both $X$ and $Y$ explained by each IV is less than $1/2$, which is generally true in genetics.

# Web Appendix C: Proof of Theorem 2

The following proofs hold for both cases with ($\tau_0 \neq 0$) and without ($\tau_0 = 0$) balanced horizontal pleiotropy. In fact, no horizontal pleiotropy is a special case of balanced horizontal pleiotropy with $\tau_0 = 0$.

## C.1 Bias of the dIVW Estimator with IV selection

When an independent selection dataset $\{\hat{\gamma}_j^*, \sigma_{\hat{\gamma}_j^*}\}$ $j = 1, \cdots, p$ is available, the IVs are included in the analysis with $|\hat{\gamma}_j^*|/\sigma_{\hat{\gamma}_j^*} > \lambda$. Then, the dIVW estimator with IV selection can be written as

$$\hat{\beta}_{\lambda,\text{dIVW}} = \frac{\hat{\theta}_{1,\lambda}}{\hat{\theta}_{2,\lambda}} = \frac{\sum_{j=1}^{p} \hat{\sigma}_{\hat{\Gamma}_j}^{-2} \hat{\gamma}_j \hat{\Gamma}_j s_j}{\sum_{j=1}^{p} \hat{\sigma}_{\hat{\Gamma}_j}^{-2} \left(\hat{\gamma}_j^2 - \hat{\sigma}_{\hat{\gamma}_j}^2\right) s_j},$$

where $s_j = I\left(|\hat{\gamma}_j^*|/\sigma_{\hat{\gamma}_j^*} > \lambda\right)$ with the indicator function $I(\cdot)$. Similar to the proof in Subection A.1 of Appendix A, the bias of the dIVW estimator with IV selection is

$$E\left(\hat{\beta}_{\lambda,\text{dIVW}} - \beta_0\right) = \beta_0 \left(\frac{v_{2,\lambda}}{\theta_{2,\lambda}^2} - \frac{v_{12,\lambda}}{\theta_{1,\lambda}\theta_{2,\lambda}}\right) + o(\varphi_\lambda) = O(\varphi_\lambda),$$

where $\theta_{1,\lambda} = E(\hat{\theta}_{1,\lambda}) = \beta_0 \sum_{j=1}^{p} \sigma_{\hat{\Gamma}_j}^{-2} \gamma_j^2 q_{\lambda,j} = \Theta(\kappa_\lambda p_\lambda)$, $\theta_{2,\lambda} = E(\hat{\theta}_{2,\lambda}) = \sum_{j=1}^{p} \sigma_{\hat{\Gamma}_j}^{-2} \gamma_j^2 q_{\lambda,j} = \Theta(\kappa_\lambda p_\lambda)$, $v_{2,\lambda} = \text{Var}(\hat{\theta}_{2,\lambda}) = \sum_{j=1}^{p} \sigma_{\hat{\Gamma}_j}^{-4} \left(4\gamma_j^2 + 2\sigma_{\hat{\gamma}_j}^4\right) q_{\lambda,j} + \sum_{j=1}^{p} \sigma_{\hat{\Gamma}_j}^{-4} \gamma_j^4 q_{\lambda,j}(1 - q_{\lambda,j}) = \Theta(\kappa_\lambda p_\lambda + p_\lambda) + \Theta(\omega^2 p_\lambda)$, and $v_{12,\lambda} = \text{Cov}(\hat{\theta}_{1,\lambda}, \hat{\theta}_{2,\lambda}) = 2\beta_0 \sum_{j=1}^{p} \sigma_{\hat{\Gamma}_j}^{-4} \sigma_{\hat{\gamma}_j}^2 \gamma_j^2 q_{\lambda,j} + \beta_0 \sum_{j=1}^{p} \sigma_{\hat{\Gamma}_j}^{-4} \gamma_j^4 q_{\lambda,j}(1 - q_{\lambda,j}) = \Theta(\kappa_\lambda p_\lambda + \varphi^2 p_\lambda)$ under Assumptions 2 and 3. As the effective sample size $\psi_\lambda = \kappa_\lambda \sqrt{p_\lambda}/\max(1,\omega) \to \infty$, $\varphi_\lambda = 1/\kappa_\lambda p_\lambda + \max(1,\omega^2)/\kappa_\lambda^2 p_\lambda \to 0$.

## C.2 Bias of the mdIVW Estimator with IV selection

The mdIVW estimator with IV selection can be written as

$$\hat{\beta}_{\lambda,\text{mdIVW}} = \frac{\hat{\theta}_{1,\lambda}}{\hat{\theta}_{2,\lambda}} \left(1 - \frac{\hat{v}_{2,\lambda}}{\hat{\theta}_{2,\lambda}^2}\right) + \frac{\hat{v}_{12,\lambda}}{\hat{\theta}_{2,\lambda}^2},$$

where $\widehat{v}_{12,\lambda} = 2\sum_{j=1}^{p} \widehat{\sigma}_{\widehat{\Gamma}_j}^{-4} \widehat{\sigma}_{\widehat{\gamma}_j}^{2} \widehat{\gamma}_j \widehat{\Gamma}_j s_{\lambda,j}$ and $\widehat{v}_{2,\lambda} = \sum_{j=1}^{p} \widehat{\sigma}_{\widehat{\Gamma}_j}^{-4} \left(4\widehat{\sigma}_{\widehat{\gamma}_j}^{2} \widehat{\gamma}_j^{2} - 2\widehat{\sigma}_{\widehat{\gamma}_j}^{4}\right) s_{\lambda,j}$ are the estimates of $v_{12,\lambda}$ and $v_{2,\lambda}$, respectively.

Let $\widehat{x}_{1,\lambda} = \widehat{\theta}_{1,\lambda}/\theta_{1,\lambda} - 1$, and $\widehat{x}_{2,\lambda} = \widehat{\theta}_{2,\lambda}/\theta_{2,\lambda} - 1$. Similar to the proof in Subsection A.2 of Appendix A, the bias of the mdIVW estimator with IV selection is

$$E\left(\widehat{\beta}_{\lambda,\mathrm{mdIVW}} - \beta_0\right) = \beta_0 \underbrace{E\left(\frac{v_{2,\lambda}}{\theta_{2,\lambda}^{2}} - \frac{\widehat{v}_{2,\lambda}}{\widehat{\theta}_{2,\lambda}^{2}}\right)}_{\mathbf{A}_\lambda} + \underbrace{E\left(\frac{\widehat{v}_{12,\lambda}}{\widehat{\theta}_{2,\lambda}^{2}} - \frac{v_{12,\lambda}}{\theta_{2,\lambda}^{2}}\right)}_{\mathbf{B}_\lambda} + \underbrace{E\left(-\widehat{x}_{1,\lambda}\frac{\widehat{v}_{2,\lambda}}{\widehat{\theta}_{2,\lambda}^{2}} + \widehat{x}_{2,\lambda}\frac{\widehat{v}_{2,\lambda}}{\widehat{\theta}_{2,\lambda}^{2}}\right)}_{\mathbf{C}_\lambda} + O\left(\varphi_\lambda^2\right).$$

For $\mathbf{A}_\lambda$, let $\widehat{y}_{2,\lambda} = \widehat{v}_{2,\lambda}/v_{2,\lambda}^{*} - 1$, where $v_{2,\lambda}^{*} = E\left(\widehat{v}_{2,\lambda}\right) = \sum_{j=1}^{p} \sigma_{\widehat{\Gamma}_j}^{-4} \left(4\sigma_{\widehat{\gamma}_j}^{2} \gamma_j^{2} + 2\sigma_{\widehat{\gamma}_j}^{4}\right) q_{\lambda,j} = \Theta\left(k_\lambda p_\lambda + p_\lambda\right)$. Through Taylor series expansion we have

$$\frac{\widehat{v}_{2,\lambda}}{\widehat{\theta}_{2,\lambda}^{2}} = \frac{v_{2,\lambda}^{*}(1+\widehat{y}_{2,\lambda})}{\theta_{2,\lambda}^{2}(\widehat{x}_{2,\lambda}+1)^{2}} = \frac{v_{2,\lambda}^{*}}{\theta_{2,\lambda}^{2}}\left\{1 - 2\widehat{x}_{2,\lambda} + 3\widehat{x}_{2,\lambda}^{2} + \widehat{y}_{2,\lambda} - 2\widehat{x}_{2,\lambda}\widehat{y}_{2,\lambda} + o_p(\varphi_\lambda)\right\}.$$

Then we have

$$\beta_0 E\left(\frac{v_{2,\lambda}}{\theta_{2,\lambda}^{2}} - \frac{\widehat{v}_{2,\lambda}}{\widehat{\theta}_{2,\lambda}^{2}}\right) = \frac{\beta_0 \sum_{j=1}^{p} \sigma_{\widehat{\Gamma}_j}^{-4} \gamma_j^{4} q_{\lambda,j}(1-q_{\lambda,j})}{\theta_{2,\lambda}^{2}} + \frac{\beta_0 v_{2,\lambda}^{*}}{\theta_{2,\lambda}^{2}}\left(\frac{2\mathrm{Cov}\left(\widehat{\theta}_{2,\lambda}, \widehat{v}_{2,\lambda}\right)}{\theta_{2,\lambda} v_{2,\lambda}^{*}} - \frac{3v_{2,\lambda}}{\theta_{2,\lambda}^{2}} + o(\varphi_\lambda)\right)$$

$$= \frac{\beta_0 \sum_{j=1}^{p} \sigma_{\widehat{\Gamma}_j}^{-4} \gamma_j^{4} q_{\lambda,j}(1-q_{\lambda,j})}{\theta_{2,\lambda}^{2}} + O\left(\varphi_\lambda^{2}\right), \qquad (11)$$

where $\mathrm{Cov}(\widehat{\theta}_{2,\lambda}, \widehat{v}_{2,\lambda}) = 8\sum_{j=1}^{p} \sigma_{\widehat{\Gamma}_j}^{-6}\sigma_{\widehat{\gamma}_j}^{4}\left(\sigma_{\widehat{\gamma}_j}^{2} + 2\gamma_j^{2}\right)q_{\lambda,j} + \sum_{j=1}^{p} \sigma_{\widehat{\Gamma}_j}^{-6}\sigma_{\widehat{\gamma}_j}^{2}\left(2\gamma_j^{2}\sigma_{\widehat{\gamma}_j}^{2} + 4\gamma_j^{4}\right)q_{\lambda,j}(1-q_{\lambda,j}) = \Theta\left(k_\lambda p_\lambda + p_\lambda\right) + \Theta(\omega^2 p_\lambda)$.

For $\mathbf{B}_\lambda$, let $\widehat{y}_{12,\lambda} = \widehat{v}_{12,\lambda}/v_{12,\lambda}^{*} - 1$, where $v_{12,\lambda}^{*} = E\left(\widehat{v}_{12,\lambda}\right) = 2\beta_0 \sum_{j=1}^{p} \sigma_{\widehat{\Gamma}_j}^{-4}\sigma_{\widehat{\gamma}_j}^{2} \gamma_j^{2} q_{\lambda,j} = \Theta\left(k_\lambda p_\lambda\right)$. Through Taylor series expansion we have

$$\frac{\widehat{v}_{12,\lambda}}{\widehat{\theta}_{2,\lambda}^{2}} = \frac{v_{12,\lambda}^{*}(1+\widehat{y}_{12,\lambda})}{\theta_{2,\lambda}^{2}(\widehat{x}_{2,\lambda}+1)^{2}} = \frac{v_{12,\lambda}^{*}}{\theta_{2,\lambda}^{2}}\left(1 - 2\widehat{x}_{2,\lambda} + 3\widehat{x}_{2,\lambda}^{2} + \widehat{y}_{12,\lambda} - 2\widehat{x}_{2,\lambda}\widehat{y}_{12,\lambda} + o_p(\varphi_\lambda)\right).$$

Then we have

$$E\left(\frac{\widehat{v}_{12,\lambda}}{\widehat{\theta}_{2,\lambda}^{2}} - \frac{v_{12,\lambda}}{\theta_{2,\lambda}^{2}}\right) = -\frac{\beta_0 \sum_{j=1}^{p} \sigma_{\widehat{\Gamma}_j}^{-4} \gamma_j^{4} q_{\lambda,j}(1-q_{\lambda,j})}{\theta_{2,\lambda}^{2}} + \frac{v_{12,\lambda}^{*}}{\theta_{2,\lambda}^{2}}\left(\frac{3v_{2,\lambda}}{\theta_{2,\lambda}^{2}} - \frac{2\mathrm{Cov}\left(\widehat{\theta}_{2,\lambda}, \widehat{v}_{12,\lambda}\right)}{\theta_{2,\lambda} v_{12,\lambda}^{*}} + o(\varphi_\lambda)\right)$$

$$= -\frac{\beta_0 \sum_{j=1}^{p} \sigma_{\widehat{\Gamma}_j}^{-4} \gamma_j^{4} q_{\lambda,j}(1-q_{\lambda,j})}{\theta_{2,\lambda}^{2}} + O\left(\varphi_\lambda^{2}\right), \qquad (12)$$

11

where $\text{Cov}(\widehat{\theta}_{2,\lambda}, \widehat{v}_{12,\lambda}) = 4\beta_0 \sum_{j=1}^{p} \sigma_{\widehat{\Gamma}_j}^{-6} \sigma_{\widehat{\gamma}_j}^{4} \gamma_j^2 q_{\lambda,j} + 2\beta_0 \sum_{j=1}^{p} \sigma_{\widehat{\Gamma}_j}^{-6} \sigma_{\widehat{\gamma}_j}^{2} \gamma_j^4 q_{\lambda,j}(1-q_{\lambda,j}) = \Theta(k_\lambda p_\lambda + \omega^2 p_\lambda)$.

For $\mathbf{C}_\lambda$, through Taylor series expansion we have

$$\widehat{x}_{1,\lambda} \frac{\widehat{v}_{2,\lambda}}{\widehat{\theta}_{2,\lambda}^2} = \widehat{x}_{1,\lambda} \frac{v_{2,\lambda}^*(1+\widehat{y}_{2,\lambda})}{\theta_{2,\lambda}^2(\widehat{x}_{2,\lambda}+1)^2} = \frac{v_{2,\lambda}^*}{\theta_{2,\lambda}^2}\{\widehat{x}_{1,\lambda} - 2\widehat{x}_{1,\lambda}\widehat{x}_{2,\lambda} + \widehat{x}_{1,\lambda}\widehat{y}_{2,\lambda} + o_p(\varphi_\lambda)\},$$

$$\widehat{x}_{2,\lambda} \frac{\widehat{v}_{2,\lambda}}{\widehat{\theta}_{2,\lambda}^2} = \widehat{x}_{2,\lambda} \frac{v_{2,\lambda}^*(1+\widehat{y}_{2,\lambda})}{\theta_{2,\lambda}^2(\widehat{x}_{2,\lambda}+1)^2} = \frac{v_{2,\lambda}^*}{\theta_{2,\lambda}^2}\{\widehat{x}_{2,\lambda} - 2\widehat{x}_{2,\lambda}^2 + \widehat{x}_{2,\lambda}\widehat{y}_{2,\lambda} + o_p(\varphi_\lambda)\}.$$

Then we have

$$E\left(-\widehat{x}_{1,\lambda}\frac{\widehat{v}_{2,\lambda}}{\widehat{\theta}_{2,\lambda}^2} + \widehat{x}_{2,\lambda}\frac{\widehat{v}_{2,\lambda}}{\widehat{\theta}_{2,\lambda}^2}\right) = \frac{v_{2,\lambda}^*}{\theta_{2,\lambda}^2}\left(\frac{2v_{12,\lambda}}{\theta_{1,\lambda}\theta_{2,\lambda}} - \frac{\text{Cov}(\widehat{\theta}_{1,\lambda}, \widehat{v}_{2,\lambda})}{\theta_{1,\lambda}v_{2,\lambda}^*} - \frac{2v_{2,\lambda}}{\theta_{2,\lambda}^2} + \frac{\text{Cov}(\widehat{\theta}_{2,\lambda}, \widehat{v}_{12,\lambda})}{\theta_{2,\lambda}v_{12,\lambda}^*} + o(\varphi_\lambda)\right)$$
$$= O(\varphi_\lambda^2), \tag{13}$$

where $\text{Cov}(\widehat{\theta}_{1,\lambda}, \widehat{v}_{2,\lambda}) = 8\beta_0 \sum_{j=1}^{p} \sigma_{\widehat{\Gamma}_j}^{-6} \sigma_{\widehat{\gamma}_j}^{4} \gamma_j^2 q_{\lambda,j} + \beta_0 \sum_{j=1}^{p} \sigma_{\widehat{\Gamma}_j}^{-6} \sigma_{\widehat{\gamma}_j}^{2} \left(4\gamma_j^4 + 2\sigma_{\widehat{\gamma}_j}^2 \gamma_j^2\right) q_{\lambda,j}(1-q_{\lambda,j}) = \Theta(k_\lambda p_\lambda + \varphi_\lambda^2 p_\lambda)$. Combining equations (11-13), we have

$$E\left(\widehat{\beta}_{\lambda,\text{mdIVW}} - \beta_0\right) = O(\varphi_\lambda^2).$$

## C.3 the Variance Difference between $\widehat{\beta}_{\lambda,\text{dIVW}}$ and $\widehat{\beta}_{\lambda,\text{mdIVW}}$

Since $\widehat{\beta}_{\lambda,\text{mdIVW}} = \widehat{\beta}_{\lambda,\text{dIVW}} + \frac{\widehat{v}_{12,\lambda}}{\widehat{\theta}_{2,\lambda}^2} - \frac{\widehat{\theta}_{1,\lambda}\widehat{v}_{2,\lambda}}{\widehat{\theta}_{2,\lambda}^3}$, we have

$$\text{Var}\left(\widehat{\beta}_{\lambda,\text{mdIVW}}\right) = \text{Var}\left(\widehat{\beta}_{\lambda,\text{dIVW}}\right) + \text{Var}\left(\frac{\widehat{v}_{12,\lambda}}{\widehat{\theta}_{2,\lambda}^2} - \frac{\widehat{\theta}_{1,\lambda}\widehat{v}_{2,\lambda}}{\widehat{\theta}_{2,\lambda}^3}\right) + 2\text{Cov}\left(\widehat{\beta}_{\lambda,\text{dIVW}} - \beta_0, \frac{\widehat{v}_{12,\lambda}}{\widehat{\theta}_{2,\lambda}^2} - \frac{\widehat{\theta}_{1,\lambda}\widehat{v}_{2,\lambda}}{\widehat{\theta}_{2,\lambda}^3}\right)$$
$$= \text{Var}\left(\widehat{\beta}_{\lambda,\text{dIVW}}\right) + 2\text{Cov}\left(\widehat{\beta}_{\lambda,dIVW} - \beta_0, \frac{\widehat{v}_{12,\lambda}}{\widehat{\theta}_{2,\lambda}^2} - \frac{\widehat{\theta}_{1,\lambda}\widehat{v}_{2,\lambda}}{\widehat{\theta}_{2,\lambda}^3}\right) + o(\varphi_\lambda^2)$$

By Taylor series expansion, we have

$$\widehat{\beta}_{\lambda,dIVW} - \beta_0 = \beta_0(\widehat{x}_{1,\lambda} - \widehat{x}_{2,\lambda}) + O_p(\varphi_\lambda),$$

$$\frac{\widehat{v}_{12,\lambda}}{\widehat{\theta}_{2,\lambda}^2} - \frac{\widehat{\theta}_{1,\lambda}\widehat{v}_{2,\lambda}}{\widehat{\theta}_{2,\lambda}^3} = \left(\frac{v_{12,\lambda}^*}{\theta_{2,\lambda}^2} - \frac{\theta_{1,\lambda}v_{2,\lambda}^*}{\theta_{2,\lambda}^3}\right) - \frac{\theta_{1,\lambda}v_{2,\lambda}^*}{\theta_{2,\lambda}^3}\widehat{x}_{1,\lambda} + \left(\frac{3\theta_{1,\lambda}v_{2,\lambda}^*}{\theta_{2,\lambda}^3} - \frac{2v_{12,\lambda}^*}{\theta_{2,\lambda}^2}\right)\widehat{x}_{2,\lambda} + \frac{v_{12,\lambda}^*}{\theta_{2,\lambda}^2}\widehat{y}_{12,\lambda}$$
$$- \frac{\theta_{1,\lambda}v_{2,\lambda}^*}{\theta_{2,\lambda}^3}\widehat{y}_{2,\lambda} + O_p(\varphi_\lambda^2).$$

12

Therefore, we have

$$\mathrm{Cov}\left(\widehat{\beta}_{\lambda,dIVW} - \beta_0, \frac{\widehat{v}_{12,\lambda}}{\widehat{\theta}_{2,\lambda}^2} - \frac{\widehat{\theta}_{1,\lambda}\widehat{v}_{2,\lambda}}{\widehat{\theta}_{2,\lambda}^3}\right)$$

$$=\mathrm{Cov}\left\{\beta_0\left(\widehat{x}_{1,\lambda} - \widehat{x}_{2,\lambda}\right), -\frac{\theta_{1,\lambda}v_{2,\lambda}^*}{\theta_{2,\lambda}^3}\widehat{x}_{1,\lambda} + \left(\frac{3\theta_{1,\lambda}v_{2,\lambda}^*}{\theta_{2,\lambda}^3} - \frac{2v_{12,\lambda}^*}{\theta_{2,\lambda}^2}\right)\widehat{x}_{2,\lambda} + \frac{v_{12,\lambda}^*}{\theta_{2,\lambda}^2}\widehat{y}_{12,\lambda} - \frac{\theta_{1,\lambda}v_{2,\lambda}^*}{\theta_{2,\lambda}^3}\widehat{y}_{2,\lambda}\right\} + o\left(\varphi_\lambda^2\right)$$

$$=\beta_0\left\{-\frac{\theta_{1,\lambda}v_{2,\lambda}^*}{\theta_{2,\lambda}^3}\mathrm{Var}\left(\widehat{x}_{1,\lambda}\right) + \left(\frac{4\theta_{1,\lambda}v_{2,\lambda}^*}{\theta_{2,\lambda}^3} - \frac{2v_{12,\lambda}^*}{\theta_{2,\lambda}^2}\right)\mathrm{Cov}\left(\widehat{x}_{1,\lambda},\widehat{x}_{2,\lambda}\right) - \left(\frac{3\theta_{1,\lambda}v_{2,\lambda}^*}{\theta_{2,\lambda}^3} - \frac{2v_{12,\lambda}^*}{\theta_{2,\lambda}^2}\right)\mathrm{Var}\left(\widehat{x}_{2,\lambda}\right)\right.$$

$$\left. + \frac{v_{12,\lambda}^*}{\theta_{2,\lambda}^2}\mathrm{Cov}\left(\widehat{x}_{1,\lambda},\widehat{y}_{12,\lambda}\right) - \frac{\theta_{1,\lambda}v_{2,\lambda}^*}{\theta_{2,\lambda}^3}\mathrm{Cov}\left(\widehat{x}_{1,\lambda},\widehat{y}_{2,\lambda}\right) - \frac{v_{12,\lambda}^*}{\theta_{2,\lambda}^2}\mathrm{Cov}\left(\widehat{x}_{2,\lambda},\widehat{y}_{12,\lambda}\right) + \frac{\theta_{1,\lambda}v_{2,\lambda}^*}{\theta_{2,\lambda}^3}\mathrm{Cov}\left(\widehat{x}_{2,\lambda},\widehat{y}_{2,\lambda}\right)\right\}$$

$$+ o\left(\varphi_\lambda^2\right)$$

$$= -\frac{\beta_0^2}{\theta_{2,\lambda}^4}\left\{\frac{v_{1,\lambda}v_{2,\lambda}^*}{\beta_0^2} - \frac{4v_{12,\lambda}v_{2,\lambda}^*}{\beta_0^2} - \frac{2v_{2,\lambda}v_{12,\lambda}^*}{\beta_0^2} + \frac{2v_{12,\lambda}v_{12,\lambda}^*}{\beta_0^2} + 3v_{2,\lambda}v_{2,\lambda}^* - \frac{\theta_{2,\lambda}\mathrm{Cov}\left(\widehat{\theta}_{1,\lambda},\widehat{v}_{12,\lambda}\right)}{\beta_0^2}\right.$$

$$\left. + \frac{\theta_{2,\lambda}\mathrm{Cov}\left(\widehat{\theta}_{1,\lambda},\widehat{v}_{2,\lambda}\right)}{\beta_0} + \frac{\theta_{2,\lambda}\mathrm{Cov}\left(\widehat{\theta}_{2,\lambda},\widehat{v}_{12,\lambda}\right)}{\beta_0} - \theta_{2,\lambda}\mathrm{Cov}\left(\widehat{\theta}_{2,\lambda},\widehat{v}_{2,\lambda}\right)\right\} + o\left(\varphi_\lambda^2\right)$$

$$= -\frac{\beta_0^2}{\theta_{2,\lambda}^4}\Delta_\lambda + o\left(\varphi_\lambda^2\right),$$

where

$$\mathrm{Cov}\left(\widehat{\theta}_{1,\lambda},\widehat{v}_{12,\lambda}\right) = 2\sum_{j=1}^{p}\sigma_{\widehat{\Gamma}_j}^{-6}\sigma_{\widehat{\gamma}_j}^2\left\{\beta_0^2\sigma_{\widehat{\gamma}_j}^2\gamma_j^2 + \left(\sigma_{\widehat{\Gamma}_j}^2 + \tau_0^2\right)\left(\gamma_j^2 + \sigma_{\widehat{\gamma}_j}^2\right)\right\}q_{\lambda,j} + 2\beta_0^2\sum_{j=1}^{p}\sigma_{\widehat{\Gamma}_j}^{-6}\sigma_{\widehat{\gamma}_j}^2\gamma_j^4 q_{\lambda,j}\left(1 - q_{\lambda,j}\right)$$

$$= \Theta\left(k_\lambda p_\lambda + p_\lambda\right) + \Theta(\omega^2 p_\lambda),$$

$$v_{1,\lambda} = \mathrm{Var}\left(\widehat{\theta}_{1,\lambda}\right) = \sum_{j=1}^{p}\sigma_{\widehat{\Gamma}_j}^{-4}\left\{\beta_0^2\sigma_{\widehat{\gamma}_j}^2\gamma_j^2 + \left(\sigma_{\widehat{\Gamma}_j}^2 + \tau_0^2\right)\left(\gamma_j^2 + \sigma_{\widehat{\gamma}_j}^2\right)\right\}q_{\lambda,j} + \beta_0^2\sum_{j=1}^{p}\sigma_{\widehat{\Gamma}_j}^{-4}\gamma_j^4 q_{\lambda,j}\left(1 - q_{\lambda,j}\right)$$

$$= \Theta\left(k_\lambda p_\lambda + p_\lambda\right) + \Theta(\omega^2 p_\lambda),$$

and

$$\Delta_\lambda = \frac{v_{1,\lambda}v_{2,\lambda}^*}{\beta_0^2} - \frac{4v_{12,\lambda}v_{2,\lambda}^*}{\beta_0} - \frac{2v_{12,\lambda}^*v_{2,\lambda}^*}{\beta_0} + \frac{2v_{12,\lambda}^{*2}}{\beta_0^2} + 3v_{2,\lambda}v_{2,\lambda}^* - \theta_{2,\lambda}\sum_{j=1}^{p}\sigma_{\widehat{\Gamma}_j}^{-6}\sigma_{\widehat{\gamma}_j}^6\left(6\sigma_{\widehat{\gamma}_j}^{-2}\gamma_j^2 + 8\right)q_{\lambda,j}$$

$$- 2\theta_{2,\lambda}\beta_0^{-2}\sum_{j=1}^{p}\sigma_{\widehat{\Gamma}_j}^{-4}\sigma_{\widehat{\gamma}_j}^4\left(\sigma_{\widehat{\gamma}_j}^{-2}\gamma_j^2 + 1\right)\left(\sigma_{\widehat{\Gamma}_j}^{-2}\tau_0^2 + 1\right)q_{\lambda,j}.$$

So, we have

$$\mathrm{Var}\left(\widehat{\beta}_{\lambda,\mathrm{dIVW}}\right) - \mathrm{Var}\left(\widehat{\beta}_{\lambda,\mathrm{mdIVW}}\right) = \frac{2\beta_0^2}{\theta_{2,\lambda}^4}\Delta_\lambda + o\left(\varphi_\lambda^2\right).$$

# Web Appendix D: Proof of $\Delta_\lambda > 0$

The following proofs hold for both cases with ($\tau_0 \neq 0$) and without ($\tau_0 = 0$) balanced horizontal pleiotropy. No horizontal pleiotropy is a special case of balanced horizontal pleiotropy with $\tau_0 = 0$.

After some algebra we have

$$\Delta_\lambda = \frac{v_{1,\lambda} v_{2,\lambda}^*}{\beta_0^2} - \frac{4 v_{12,\lambda} v_{2,\lambda}^*}{\beta_0} - \frac{2 v_{12,\lambda}^* v_{2,\lambda}^*}{\beta_0} + \frac{2 v_{12,\lambda}^* v_{12,\lambda}^*}{\beta_0^2} + 3 v_{2,\lambda} v_{2,\lambda}^* - \theta_2 \sum_{j=1}^p \sigma_{\hat{\Gamma}_j}^{-6} \sigma_{\hat{\gamma}_j}^6 (8 + 6 k_j) q_{\lambda,j}$$

$$- 2 \beta_0^{-2} \theta_{2,\lambda} \sum_{j=1}^p \left( \sigma_{\hat{\Gamma}_j}^{-4} \sigma_{\hat{\gamma}_j}^4 + \sigma_{\hat{\Gamma}_j}^{-6} \sigma_{\hat{\gamma}_j}^4 \tau_0^2 \right) (1 + k_j) q_{\lambda,j}$$

$$= \left\{ v_{2,\lambda}^* \sum_{j=1}^p \left\{ \sigma_{\hat{\Gamma}_j}^{-4} \sigma_{\hat{\gamma}_j}^4 k_j + \beta_0^{-2} \sigma_{\hat{\Gamma}_j}^{-2} \sigma_{\hat{\gamma}_j}^2 (1 + k_j) \right\} q_{\lambda,j} + 6 v_{2,\lambda}^* \sum_{j=1}^p \sigma_{\hat{\Gamma}_j}^{-4} \sigma_{\hat{\gamma}_j}^4 q_{\lambda,j} + 8 \left( \sum_{j=1}^p \sigma_{\hat{\Gamma}_j}^{-4} \sigma_{\hat{\gamma}_j}^4 k_j q_{\lambda,j} \right)^2 \right.$$

$$\left. - \theta_{2,\lambda} \sum_{j=1}^p \sigma_{\hat{\Gamma}_j}^{-6} \sigma_{\hat{\gamma}_j}^6 (8 + 6 k_j) q_{\lambda,j} - 2 \beta_0^{-2} \theta_{2,\lambda} \sum_{j=1}^p \sigma_{\hat{\Gamma}_j}^{-4} \sigma_{\hat{\gamma}_j}^4 (1 + k_j) q_{\lambda,j} \right\}$$

$$+ \beta_0^{-2} \tau_0^2 \left\{ v_{2,\lambda}^* \sum_{j=1}^p \sigma_{\hat{\Gamma}_j}^{-4} \sigma_{\hat{\gamma}_j}^2 (1 + k_j) q_{\lambda,j} - 2 \theta_{2,\lambda} \sum_{j=1}^p \sigma_{\hat{\Gamma}_j}^{-6} \sigma_{\hat{\gamma}_j}^4 (1 + k_j) q_{\lambda,j} \right\}$$

$$= \mathbf{D}_\lambda + \mathbf{E}_\lambda,$$

where $\mathbf{D}_\lambda$ represents the part of the curly brackets before the plus sign, and $\mathbf{E}_\lambda$ represents the part after the plus sign. A sufficient condition for $\Delta_\lambda > 0$ is $\mathbf{D}_\lambda > 0$ and $\mathbf{E}_\lambda \geqslant 0$, which is similar to the proof of $\mathbf{D} > 0$ and $\mathbf{E} \geqslant 0$ in Appendix B, so we omit the detials here. And the condition for $\Delta_\lambda > 0$ is that

$$\max(h_{X,j}, h_{Y,j}) < 1/2.$$

# Web Tables

**Web Table 1** Comparison of the mdIVW estimator with other competing MR methods. The ture causal effect $\beta_0 = 0.5$. Balanced horizontal pleiotropy exists ($\tau_0 = 0.01$) and no IV selection is conducted. The simulation is based on 20,000 repetitions. Bias(%): bias divided by $\beta_0$; SE: average of standard errors; CP: coverage probability of the 95% confidence interval.

| $\psi$ | Method | Bias(%) | SE | CP |
|---|---|---|---|---|
| 8.14 | IVW | -79.49 | 0.063 | 0.000 |
| | MR-Median | -63.37 | 0.094 | 0.082 |
| | MR-Egger | -65.00 | 0.097 | 0.085 |
| | MR-RAPS | 9.71 | 0.512 | 0.951 |
| | dIVW | 4.55 | 0.345 | 0.962 |
| | mdIVW | -0.50 | 0.319 | 0.953 |
| 17.83 | IVW | -63.86 | 0.057 | 0.000 |
| | MR-Median | -44.00 | 0.086 | 0.278 |
| | MR-Egger | -43.83 | 0.083 | 0.251 |
| | MR-RAPS | 1.26 | 0.170 | 0.949 |
| | dIVW | 1.21 | 0.165 | 0.953 |
| | mdIVW | 0.17 | 0.163 | 0.952 |
| 27.90 | IVW | -53.14 | 0.052 | 0.001 |
| | MR-Median | -35.12 | 0.077 | 0.369 |
| | MR-Egger | -32.04 | 0.075 | 0.426 |
| | MR-RAPS | 0.38 | 0.119 | 0.950 |
| | dIVW | 0.40 | 0.115 | 0.950 |
| | mdIVW | -0.09 | 0.114 | 0.950 |

**Web Table 2** Comparison of the mdIVW estimator with other competing MR methods. The true causal effect $\beta_0 = 0.5$. Balanced horizontal pleiotropy exists ($\tau_0 = 0.01$). The IV selection threshold $\lambda = \sqrt{2\log p}$. The simulation is based on $20,000$ repetitions. Bias(%): bias divided by $\beta_0$; SE: average of standard errors; CP: coverage probability of the 95% confidence interval.

| $\psi_\lambda$ | Method | Bias(%) | SE | CP |
|---|---|---|---|---|
| 7.22 | IVW | -2.97 | 0.122 | 0.934 |
| | MR-Median | -4.76 | 0.143 | 0.939 |
| | MR-Egger | -22.53 | 0.385 | 0.925 |
| | MR-RAPS | -0.08 | 0.130 | 0.941 |
| | dIVW | 0.34 | 0.127 | 0.952 |
| | mdIVW | -0.27 | 0.126 | 0.952 |
| 10.19 | IVW | -2.99 | 0.107 | 0.935 |
| | MR-Median | -5.02 | 0.134 | 0.937 |
| | MR-Egger | -25.06 | 0.316 | 0.916 |
| | MR-RAPS | 0.42 | 0.114 | 0.942 |
| | dIVW | 0.74 | 0.112 | 0.950 |
| | mdIVW | 0.26 | 0.111 | 0.951 |
| 16.02 | IVW | -4.01 | 0.092 | 0.936 |
| | MR-Median | -6.32 | 0.116 | 0.940 |
| | MR-Egger | -27.98 | 0.252 | 0.900 |
| | MR-RAPS | 0.19 | 0.099 | 0.941 |
| | dIVW | 0.37 | 0.097 | 0.953 |
| | mdIVW | 0.02 | 0.097 | 0.953 |

**Web Table 3** Description of the GWAS datasets used in this paper

| Trait | Dataset | Source | Population | Sample size (case/control) | Trait description |
|---|---|---|---|---|---|
| CAD | Selection | GERA | European | 16,399/45,448 | ICD9: 398, 402, 404, 410-414, 425-428, 430-438, 441 |
|  | Exposure | UK BioBank | European | 14,510/97,828 | ICD10: I01, I09, I11, I13, I20-I25, I42-I50, I60-I69, I71 |
| Dyslipidemia | Selection | GERA | European | 133,024/28,823 | ICD-9: 272 |
|  | Exposure | UK BioBank | European | 16,818/95,520 | ICD-10: E78 |
| Heart Failure | Outcome | HERMES | European | 47,309/930,014 | a clinical diagnosis of HF of any aetiology with no inclusion criteria based on LV ejection fraction |

**Web Table 4** The number of IVs and the estimated effective sample sizes for coronary artery disease (CAD) and dyslipidemia

| Exposure | No IV selection | | IV selection $\lambda = \sqrt{2\log p}$ | |
|---|---|---|---|---|
| | $p$ | $\widehat{\psi}$ | $\widehat{p}_\lambda$ | $\widehat{\psi}_\lambda$ |
| CAD | 2413 | 5.05 | 122 | 4.59 |
| Dyslipidemia | 2480 | 30.90 | 165 | 30.36 |

**Web Table 5** Estimated causal effects $(\widehat{\beta})$ and estimated standard errors (SEs) of coronary artery disease (CAD) and dyslipidemia on the risk of heart failure under balanced horizontal pleiotropy.

| Exposure | Exposure | No IV selection | | IV selection with $\lambda = \sqrt{2\log p}$ | |
|---|---|---|---|---|---|
| | | $\widehat{\psi}$ | $\widehat{\beta}$ (SE) | $\widehat{\psi}_\lambda$ | $\widehat{\beta}$ (SE) |
| CAD[a] | MR-RAPS | 5.05 | 0.890 (0.297) | 4.59 | 0.862 (0.252) |
| | dIVW | | 1.086 (0.356) | | 0.997 (0.313) |
| | mdIVW | | 0.982 (0.237) | | 0.909 (0.234) |
| Dyslipidemia[b] | MR-RAPS | 30.90 | 0.187 (0.034) | 30.36 | 0.207 (0.028) |
| | dIVW | | 0.195 (0.032) | | 0.203 (0.027) |
| | mdIVW | | 0.194 (0.031) | | 0.203 (0.024) |

[a]The estimated balanced pleiotropy for CAD is $\widehat{\tau}^2 = -2.27 \times 10^{-5}$;

[b]The calculated balanced pleiotropy for Dyslipidemia is $\widehat{\tau}^2 = 3.54 \times 10^{-5}$.

# Web Figures

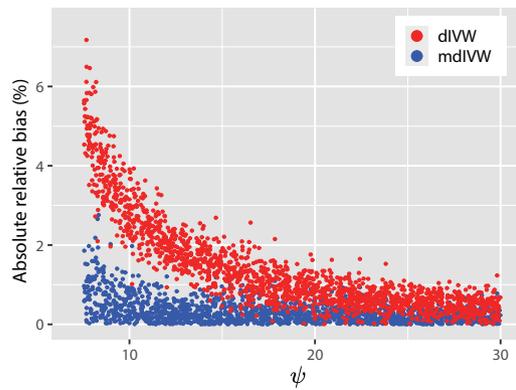

(a)

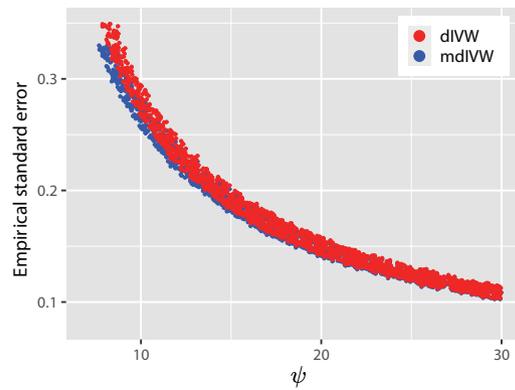

(b)

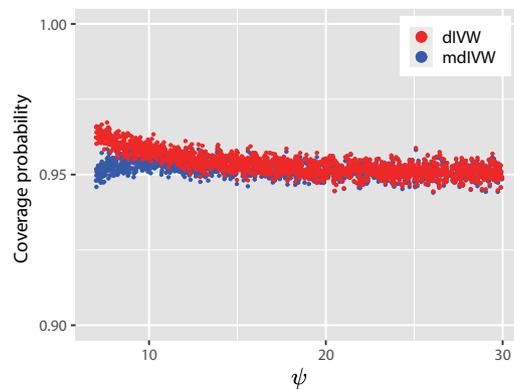

(c)

**Web Figure 1** The plots of (a) the estimated absolute relative biases (absolute biases divided by $\beta_0$); (b) the empirical standard errors; and (c) the simulation coverage probabilities of the 95% confidence intervals for the dIVW and mdIVW estimators versus the effective sample size $\psi$. The dots represent the simulation results under different combinations of settings based on 10,000 repetitions. The balanced horizontal pleiotropy exists ($\tau_0 = 0.01$). There is no IV selection.

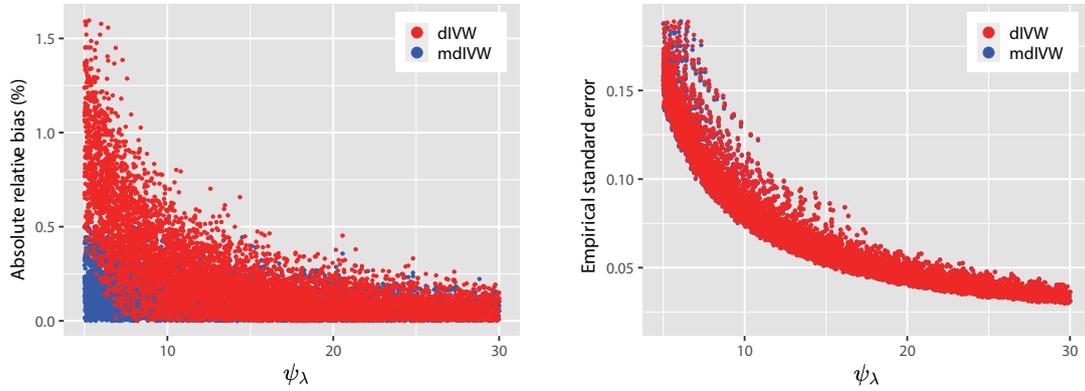

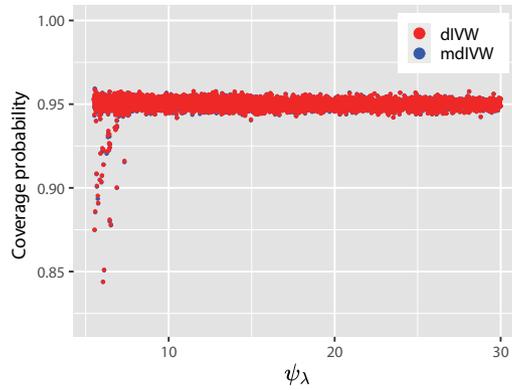

**Web Figure 2** The plots of (a) the estimated absolute relative biases (absolute biases divided by $\beta_0$); (b) the empirical standard errors; and (c) the simulation coverage probabilities of the 95% confidence intervals for the dIVW and mdIVW estimators versus the effective sample size $\psi_\lambda$. The dots represent the simulation results under different combinations of settings based on 10,000 repetitions. The balanced horizontal pleiotropy exists ($\tau_0 = 0.01$). The IV selection threshold $\lambda = \sqrt{2\log p}$.



\end{document}